\documentstyle[epsfig,12pt]{article}
\def\Journal#1#2#3#4{{#1} {#2} (#4) #3 }

\def\PPNP{\em Prog. Part. Nucl. Phys.}
\def\PLB{{\em Phys. Lett.} B}

\def\PRC{{\em Phys. Rev.} C}
\def\PRD{{\em Phys. Rev.} D}
\def\PRL{\em Phys. Rev. Lett.}

\def\PREP{\em Phys. Rep.}

\def\EPJA{{\em Eur. Phys. J.} A}

\def\ZPA{{\em Z. Phys.} A}

\def\NPA{{\em Nucl. Phys.} A}

\def\LNC{\em Lett. Nuovo Cimento}

\def\JPSJ{\em J. Phys. Soc. Japan}

\def \dd {\mbox{d\raisebox{0.75ex}{\hspace*{-0.28em}-}\hspace*{-0.08em}}}

\newcommand{\be}{\begin{equation}}
\newcommand{\ee}{\end{equation}}
\newcommand{\bee}{\begin{eqnarray}}
\newcommand{\eee}{\end{eqnarray}}

\newcommand{\piNetaN}{$\pi N \! \rightarrow \! \eta N \:$}

\begin{document}

$\phantom{.}$

\vspace{6cm}

\begin{center}
{\huge International Workshop on $\eta$--Nucleus Physics}

\vspace{2cm}

{\large May 8--12, 2006, J\"ulich, Germany}

\vspace{1cm}

{\large Summary}

\vspace{1cm}

{\large and}

\vspace{1cm}

{\huge  Working Group Progress Report}

\end{center}

\newpage

{$\phantom{=}$}

\vspace{0.5cm}

\tableofcontents

\newpage

%

\section{Introduction}


\subsection{About the network}
\addtocontents{toc}{\hspace{2cm}(B. H\"oistad)\par}

\vspace{5mm}

B. H\"oistad

\vspace{5mm}
\noindent
 Department of
Radiation Sciences of Uppsala University,
 Uppsala, Sweden

\vspace{5mm}

\noindent
This workshop is recognized as part of a European Community Integrated
Infrastructure Initiative devoted to Hadron Physics (I3HP). This
initiative contains several activities, one of them being the network
EtaMesonNet which is created to exchange information on experimental
and theoretical ongoing activities on $\eta$--physics at different European
accelerator facilities and institutes, and is thereby promoting the
infrastructure in Europe. 

 The present workshop is focused on
specific aspects of $\eta$  physics, namely the physics involved in the
$\eta$--nucleon and the $\eta$--nucleus systems. It is very important to
clarify, qualitatively as well as quantitatively, our present
knowledge in this area, and try to formulate a strategy for how
further knowledge should be gained from new advanced theoretical
developments as well as from new experimental efforts. This workshop
was mainly devoted to different theoretical issues, although a close
connection with experimental data was always present. A substantial
part of the workshop was given to discussions in connection with
different selected talks. This form of the workshop turned out to be
very fruitful and profitable.

  The workshop was held in May 8-12,
2006, at the Forschungzentrum J\"ulich, enjoying kind hospitality and
support from the IKP theory division. The financial support from the
European Community-Research Infrastructure Activity under FP6
"Structuring the European Research Area" program (Hadron Physics,
contract number RII3-CT-2004-506078), is gratefully acknowledged.

\vspace{5mm}


\newpage


\subsection{Scope of the workshop and results}
\addtocontents{toc}{\hspace{2cm}(C.Hanhart)\par}

\vspace{5mm}

C.Hanhart

\vspace{5mm}

\noindent
Institut f\"ur Kernphysik (Theorie), Forschungszentrum J\"ulich

\vspace{5mm}


\noindent
At the beginning of the Workshop a set of questions was posed to
sharpen the focus of the meeting. These were:

\begin{enumerate}
\item What is the proper treatment of the $\eta$-N system within an
    $\eta$-nucleus calculation? In particular: can consistent
    equations be formulated that allow one to study a resonance
    propagator for this system?

\item  What is the proper two body input to the few-body equations
    and what data can be used to constraint it? What do we know
    about the strength and energy dependence of the $\eta$-N
    interaction?

\item  Inelastic channels: which ones are to be included and how?
    What is the origin of $\eta$-A imaginary part, how big is it
    and how can its size be determined unambiguously? This
    issue is closely connected to the previous one.

\item  To what extent can an $\eta$-A study tell us something about the
    nature of the S11(1535)? Can we agree on some strategy to
    arrive possibly even at a model-independent answer to this
    question?

\item  What needs to be done to get a consistent description of the
    $\eta$-d and $\eta$-${}^3$He systems and how can one proceed to heavier
    systems such as $\eta$-$\alpha$ for which experimental information is
    available?

\item  Are there other reactions that could shed light on the
    $\eta$-nucleus interaction?
\end{enumerate}

Using this list as a basis, in this brief summary we try to
present to what extent those questions were answered and where
further research is necessary.

1: As long as the $\eta$-N scattering length is fixed and the
effective range is large and negative, there seems to be very
little sensitivity of the $\eta$-d scattering length to the details
of the model used for $\eta$-N interaction. Nevertheless, the sensitivity to the
value of the scattering length is large. The role of the effective
range still needs to be explored systematically.

2,3: There was agreement that the large amount of high quality
data now available for reactions with $\eta$-N final state should all
be used to constrain the two-body dynamics. The role of the
two-pion channels was identified to be sizable, but not essential
for the two--body dynamics. The role of the 2-$\pi$ channels
as well as that of relativistic kinematics in the
$\eta$--A reactions needs further systematic studies.

It became clear, that the influence of the $S_{11}(1650)$ on the $\eta$
production reactions needs to be explored further and models
should be faced with the $N S_{11}$ transition form factors as measured
at CLAS.

From the discussions it emerged that there is no conflict between
a sizable imaginary part of the $\eta$-A scattering length induced by
the pion channels, contrary to recent claims. On the other hand is
a small imaginary part a precondition for a (quasi)bound state.
The issue on the size of the imaginary part can only be solved
experimentally. Fortunately, this imaginary part can in principle be extracted
model-independently from very near-threshold $\eta$-A production
data. For the $\eta$--${}^3H$e system, the new ANKE and COSY-11 measurements might
be of sufficient quality to settle this issue.

4: This remained unclear. If there is a dynamical singularity due
to a bound state in the $\eta$-A system, this singularity will
control the close-to-threshold dynamics and there is no clear cut
connection to the properties of the $S_{11}$. However, there is still a
chance that different models for the $\eta N$ dynamics will lead to
different energy dependencies of the $\eta N \to \eta N$ transition ---
encoded especially in the effective range. Thus there might still
be a chance to get more information on the $S_{11}$ from $\eta$-A studies.

5: So far there exist microscopic calculations for $\eta$--d and $\eta$--${}^3$He.
Then there are effective models for the $\eta$-A interactions for nuclei
heavier than Carbon. At present it is not clear how to connect these
regimes.

6: It would be good to have data on the near-threshold $\eta$-A
interaction for nuclei beyond ${}^4$He. Although at present no
systematic first principle calculations are possible for those
systems it is important to investigate the trends of the low
energy $\eta$-A interaction as a function of A. We know already that
the energy dependence of $\eta$-d and $\eta$-$^3$He are quite similar,
whereas that of $\eta$-$^4$He is significantly weaker. This might be
interpreted as a more strongly bound system for $\eta$ ${}^4$He, since a
more distant pole influences the near threshold regime less. To
better understand the $\eta$-A system we would therefore need data on
somewhat heavier nuclei than $^4$He. Our current believe would be
that the binding grows with growing A --- this could be read off
from the data directly.

\vspace{0.2cm}

The webpage of the workshop is http://www.fz-juelich.de/ikp/etanucleus/

\newpage

\section{Short summary of the talks}

\subsection{
Photoproduction of $\eta$-mesons off nucleons and nuclei}
\addtocontents{toc}{\hspace{2cm}(B. Krusche)\par}
\vspace{5mm}
\noindent
B. Krusche$^{(a)}$,

\vspace{5mm}
\noindent
$^{(a)}$ Department of Physics and Astronomy, University of Basel,
Ch-4056 Basel, Switzerland
\vspace{5mm}


\noindent
The talk has summarized experimental and partly also theory work on the
photoproduction of $\eta$-mesons in view of the following topics:
\begin{itemize}
\item[1)] {Photoexcitation of the resonances on the free proton, in particular
at low energies S$_{11}$(1535) resonance, D$_{13}$(1520) resonance.}
\item[2)] {Resonance contribution to $\eta$-photoproduction off the neutron,
quasifree off neutrons bound in the deuteron, helium isotopes and coherent
from deuteron}
\item[3)] {Threshold enhancements in $\eta$ photoproduction off light nuclei
and the search for $\eta$-mesic nuclei}
\item[4)]  {Photoproduction of $\eta$-mesons off heavy nuclei in view of
$\eta$-nucleus FSI ($\eta$ - nucleon absortpion cross section) and the
in-medium properties of the S$_{11}$ resonance.}
\end{itemize}

A summary on results published before 2003 concerning points 1), 2) is given 
in \cite{Krusche_03}, a short summary on 3), 4) can be found in 
\cite{Krusche_05}. The following is a short summary of the discussed topics 
with the relevant references. There is certainly more relevant work, the
reference list reflects only the work that has been adressed in some way in
the talk.

\begin{itemize}
\item {Photo-(Electro-)production of $\eta$ off the free proton:}

\begin{itemize}
\item{Measurements of the $p(\gamma ,\eta)p$ differential and total 
cross sections:\\
First isobar fit proposing S$_{11}$ dominnance is given in \cite{Hicks_73},
update with still mostly untagged pre-1990 data is given in \cite{Tabakin_89}. 
Modern measurements with tagged photon beams: Tokyo (1988), 800 - 1000 MeV,
angles of 45$^o$, 80$^o$, 100$^o$, 110$^o$ \cite{Homma_88}; 
Bonn (1995) AMADEUS (few data below 730 MeV) \cite{Price_95}; Mainz (1995)
TAPS (energies up to 800 MeV, most precise threshold measurement) 
\cite{Krusche_95,Krusche_95a,Krusche_95b}; Grenoble (2002) GRAAL (energies 
up to 1.1 GeV) \cite{Renard_02}; Jlab (2002) CLAS (energies up to 2 GeV)
\cite{Dugger_02}, Bonn (2005) Crystal Barrel (energies up to 3 GeV)
\cite{Crede_05}.}
\item{Measurements of polarization observables:\\
Photon beam asymmetry: Grenoble (1998) GRAAL \cite{Ajaka_98}; Target asymmetry:
Bonn (1998) PHOENICS \cite{Bock_98}. Remark: Model analyses (isobar models
\cite{Tiator_99}, effective Lagrangian models \cite{Mathur_98}) find consistent
solutions for angular distributions and beam asymmetry but then fail to
reproduce the target asymmetry. It is shown in \cite{Tiator_99} that 
fits which enfore the target asymmetry result in an unexpectedly large 
and strongly energy dependent phase between S$_{11}$ and D$_{13}$
multipoles. This is an unsolved problem.  
}
\item{Messurement of electroproduction cross sections: Bonn (1995) ELAN
(only total cross section very close to photon point) \cite{Schoch_95};
Jlab (2001) CLAS $Q^2\leq 1.3$ GeV$^2$ \cite{Thompson_01}, Jlab (1999) HMS
$Q^2=$ 2.4, 3.6 GeV$^2$ \cite{Armstrong_99}.
}
\item{Model analyses of data in particular in view of S$_{11}$ resonances and
D$_{13}$(1520):\\
Extraction of S$_{11}$ parameters in particular electromagnetic helicity
coupling $A_{1/2}^p$, see summary and discussion in \cite{Krusche_05},
some of the above cited experimental papers quote numbers 
\cite{Homma_88,Krusche_95, Renard_02}. Further analysis e.g. in 
\cite{Knochlein_95,Sauermann_95,Li_95a,Benmerrouche_96,Krusche_97,Tiator_99,Saghai_01,Chiang_02}. 
Refs. \cite{Tiator_99,Chiang_02} analyse also the
polarization data and extract branching ratio for 
$D_{13}$(1520)$\rightarrow N\eta$. 
Ref. \cite{Saghai_01} claims evidence for third S$_{11}$ resonance
from analysis of GRAAL data. Was not confirmed by later data and analyses 
(see e.g. \cite{Crede_05}). 
}
\item{Model predictions for the sensitivity of the $p(\gamma, \eta\gamma')p$
reaction to the magnetic moment of the S$_{11}$(1535) resonance:
\cite{Chiang_03,Hyodo_03}.\\ Model predictions for the magnetic moment
of the S$_{11}$(1535): \cite{Chiang_03,Hyodo_03,Liu_05}.
}
\item{Theory work concerned with the nature of the S$_{11}$ resonance 
(apart from standard constituent quark models descriptions):\\
Chiral constituent quark model with hyperfine interaction due to Goldstone boson
exchange \cite{Glozman_96,Glozman_96a}. Leads to quark-diquark clustering so
that selection rules derived from quantum numbers of diqaurk favor $\eta$
decay of S$_{11}$(1535) and suppress $\eta$ decay of S$_{11}$(1650).
S$_{11}$(1535) is dynamiccaly generated ($K\Sigma$ molecular-like state)
in chiral coupled channel calculations \cite{Kaiser_95,Kaiser_97}.
}
\end{itemize}
\item{Photoproduction of $\eta$ off light nuclei in view of neutron cross
section and threshold enhancements:}
\begin{itemize}
\item{Measurements of quasi-free photoproduction off the neutron bound in the
deuteron:\\
Frascati 1969, untagged photon beam \cite{Bacci_69} found roughly 
$\sigma_n\approx \sigma_p$ in S$_{11}$ region. All modern tagged photon 
experiments agree on $\sigma_n /\sigma_p\approx$2/3 in S$_{11}$(1535) range: 
Mainz (1995) TAPS \cite{Krusche_95c}; Bonn (1997) Phoenics
\cite{Hoffmann-Rothe_97}; Mainz (2003) TAPS \cite{Weiss_03}.
Threshold enhancements are discussed in \cite{Hejny_02}.
First preliminary results indicate a strong rise of the neutron/proton ratio
at higher incident photon energies: Grenoble (2004) GRAAL \cite{Kouznetsov_04};
Bonn (2005) Crystal Barrel/TAPS \cite{Jaegle_05}. Proton/neutron cross section
ratio shows peak-like structure around invariant masses of 1.65 GeV.
Probably due to resonance with strong photo-coupling to neutron, type of
resonance not yet established.
}
\item{Measurement of coherent $\eta$ photoproduction off the deuteron:\\
Stanford 1969, untagged photon beam, only detection of recoil deuteron,
found relatively large cross sections indicating a negligible iso-vector part
\cite{Anderson_69}.
Later tagged photon beam measurements with detection of recoil deuteron and
$\eta$ meson found much smaller cross sections indicating dominant iso-vector
part. Mainz (1995) TAPS \cite{Krusche_95c}, only upper limits much below
ref. \cite{Anderson_69}. Bonn (1997) PHOENICS \cite{Hoffmann-Rothe_97}
and Mainz (2001) TAPS \cite{Weiss_01} agree for most angles, beam energies,
TAPS data lower than PHOENICS data for cm angles around 90$^o$.   
}
\item{Measurement of quasi-free and coherent photoproduction of $\eta$ off
He nuclei:\\
Quasifee of $^{4}$He (Mainz/TAPS \cite{Hejny_99}) confirmed 
$\sigma_n /\sigma_p\approx$2/3. No signal for coherent production seen
in same experiment, only upper limits. Threshold behavior discussed in
\cite{Hejny_02}. Quasifree and coherent off $^3$He (Mainz/TAPS 2004)
\cite{Pfeiffer_04,Pfeiffer_05}: coherent contribution clearly identified, 
threshold enhancement of coherent part, indication of strong FSI,
tentative indication for (quasi)bound state.
}
\item{Related results (threshold enhancements) for hadron induced reactions:\\
in particular $pp\rightarrow pp\eta$ \cite{Calen_96}, $np\rightarrow d\eta$ 
\cite{Plouin_90,Calen_98}, $pd\rightarrow\eta  ^3\mbox{He}$
\cite{Mayer_96,Bilger_02}, $\vec{d}d\rightarrow \eta ^4\mbox{He}$ \cite{Willis_97}, 
and $pd\rightarrow pd\eta$ \cite{Hibou_00}. All reactions show more or less 
pronounced threshold enhancements. However, so far there is no conclusive 
evidence that the final state interaction is strong enough to form quasi-bound 
states.  
}
\item{Experimental results for the explicit search of $\eta$-mesic nuclei:\\
Pion induced reactions on oxygen nuclei, search for kinematical peak
from two-body final states in $\pi^++^{16}O\rightarrow p+^{15}_{\eta}O$ 
\cite{Chrien_88} and in $\pi^++^{18}O\rightarrow \pi^-+^{18}_{\eta}Ne$
\cite{Johnson_93}. No evidence reported.\\
Claim of positive result from the photon induced reaction chain 
$\gamma + ^{12}C\rightarrow N+_{\eta}(A-1)\rightarrow N +\pi^+ +n+(A-2)$
for $A$=11 nuclei (carbon,beryllium) \cite{Sokol_00,Sokol_01}, not yet 
supported by other experiments.\\
Result from photoproduction on $^3$He \cite{Pfeiffer_04,Pfeiffer_05},
must be confirmed with better statistics. 
}
 \item{Model calculations of $\eta$ photoproduction off deuteron and
He nuclei:\\
Quasifree and coherent photoproduction off the deuteron and deuteron
threshold effects e.g.: 
\cite{Fix_97,Ritz_99,Fix_00,Fix_00a,Ritz_01,Sibirtsev_01,Fix_02,Sibirtsev_02,Fix_04}. 
Photoproduction off the three-nucleon system e.g.: 
\cite{Fix_02a,Fix_03,Sibirtsev_04,Hanhart_05}.
}
\item{Predictions for existence of $\eta$-mesic nuclei, analysis of $\eta N$
scattering length:\\
Early results for scattering length ($a$=0.27+$i$0.22) and possible existence 
of $\eta$-mesic nuclei with $A > 10$: \cite{Bhalerao_85,Liu_86}. Many new 
results of scattering length due to availability of more precise data
(majority finds larger values of real part but span the entire range from 
0.2 - 1.0, most cluster between 0.5 - 0.8), resulting in much discussion 
about existence of light $\eta$-mesic nuclei (in particular $^2$H, $^3$H, 
$^3$He, $^4$He): 
\cite{Ueda_91,Ueda_92, Wilkin_93,Rakityanski_95,Rakityanski_96,Green_96,Green_97,
Scoccola_98,Green_99,Shevchenko_00,Grishina_00,Garcilazo_01,Sibirtsev_02a}
and probably more.
}
\end{itemize}
\item{Photoproduction off heavy nuclei,
$\eta$-nucleus FSI and S$_{11}$ in-medium properties:}
\begin{itemize}
\item{Measurements of $\eta$-photoproduction off heavy nuclei:\\
Early untagged data from Frascati \cite{Bacci_69a}. First precise measurement
from threshold to $\approx$800 MeV Mainz (1996) TAPS \cite{Roebig_96}.
Established $A^{2/3}$ scaling of cross section, i.e. strong FSI, found
$\eta N$ absorption cross section of $\approx$30 mb, mean-free path
$\lambda\approx$ 2 fm. No significant S$_{11}$(1535) modification.
Extension to energies up to $\approx$ 1 GeV from KEK 
\cite{Yorita_00,Yamazaki_00} report at most small in-medium effects.
Most recent data from LNS Tohoku University up to 1.1 GeV 
\cite{Kinoshita_05} and from Bonn (Crystal Barrel/TAPS) up to 2 GeV,
covering entire S$_{11}$ range \cite{Mertens_06}. Some additional results and
discussion of mass dependence etc. in comparison to other meson production
channels in \cite{Krusche_04,Krusche_04a}. Seem to indicate that re-scattered
contributions from nuclear volume show stronger suppession of second
resonance bump than quasi-free surface contributions.  
}
\item{Model calculations for $\eta$-photoproduction off heavy nuclei:\\
Comparison of results from cascade Monte Carlo calculations \cite{Carrasco_93} 
or BUU calculations \cite{Hombach_95,Lehr_00,Lehr_01,Lehr_03} to data show no
need for significant in-medium modification of S$_{11}$. Predictions for
in-medium spectral function of S$_{11}$ from self-consistent calculations
show only small in-medium effects \cite{Post_04}.
}
\end{itemize}
\end{itemize}



\newpage

\subsection{ The Crystal Ball data on $\pi^0$, $2\pi^0$, and $\eta$ pion-- and photoproduction on complex targets
 }
\addtocontents{toc}{\hspace{2cm}(A.~Starostin)\par}

\vspace{5mm}
A.~Starostin

\vspace{5mm}
\noindent
University of California, Los Angeles, CA 90095-1547, USA
\vspace{5mm}

\noindent
A unique set of data on light, neutral, meson production by $\pi^-$ on
complex targets has been obtained by the Crystal Ball collaboration at
Brookhaven National Laboratory. The data include angular
distributions, distributions of kinetic energy and missing mass for
$\pi^- A \to \pi^0 X$ and $\pi^- A \to \eta X$, as well as the
invariant mass distributions and the Dalitz plots for $\pi^- A \to
2\pi^0 X$.  The data were obtained for two incident $\pi^-$ beam
momenta (408~MeV/$c$, and 750~MeV/$c$) on four different targets:
$H_2$, $C$, $Al$, and $Cu$. The analysis of the single and double pion
production reveals significant pion rescattering and absorption in
complex nuclei. We concluded that the rescattering and absorption are
mainly responsible for the observed changes in the shape of the
$2\pi^0$ invariant mass for the complex targets.  Our hydrogen data
indicate that the dominant mechanism of $2\pi^0$ production on a
proton is via the intermediate $\Delta(1232)$ state. The contribution
of the alternative mechanism via the intermediate ${\rm f_0}(600)$
meson is not significant.  The results obtained on complex targets
show a strong nuclear absorption of the $\Delta(1232)$. This may
enhance the fraction of the $2\pi^0$ events produced via the ${\rm
f_0}(600)$. See Refs~[1,2,3] for details.

New Crystal Ball/TAPS data has been obtained in 2005 at the Mainz
Microtron Facility. The experiment uses the MAMI real photon beam with
the continuous energy spectrum from 100~MeV to the maximum energy of
819~MeV. The list of the targets include $H_2$, $D_2$, $Li$, $C$,
$HO_2$, $Ca$, and $Pb$. The data were obtained for the following
reactions: $\gamma A \to \pi^0 X$, $\gamma A \to \eta X$, and $\gamma
A \to 2\pi^0 X$. Analysis of the experimental data is in progress.

\newpage

\subsection{ Investigation of the $^3$He-$\eta$ Final State in
dp-Reactions at ANKE }
\addtocontents{toc}{\hspace{2cm}(T. Mersmann)\par}

 \vspace{5mm} T. Mersmann

\vspace{5mm}
\noindent
Institut f\"ur Kernphysik, Universit\"at M\"unster, M\"unster, Germany
\vspace{5mm}

\noindent
The existence of $\eta$-mesic $^3$He-nuclei is still an open issue
of research [1,2,3]. To investigate the possibility of the
formation of such a bound system, production measurements with one
$\eta$ meson and the $^3$He-nucleus in the final state are of
great interest. By studying this system at low
excess energies, information about the final state interaction and
therefore about the scattering length of the $\eta$-nucleus system
can be gained. The latter one is closely related to the properties
of such a possible bound state and has to be determined with high
precision.

The available data sets for p$+$d$\rightarrow$$^3$He$+\eta$
production experiments in the close vicinity of the threshold
[4,5] expose discrepancies, which currently forbid the extraction
of scattering length information with sufficient precision.
Therefore, the reaction d$+$p$\rightarrow$$^3$He$+\eta$ has been
investigated at the ANKE spectrometer [6] using a continuously
ramped accelerator beam at excess energies ranging from below
threshold up to Q=+12 MeV. Due to the full geometrical acceptance
of the ANKE spectrometer high statistic data on this reaction have
been obtained. Additionally, data at excess energies of Q = 20, 40
and 60 MeV have been recorded in order to determine total cross
sections and to investigate contributions from higher partial
waves.

The identification of $^3$He-nuclei is done using scintillation
wall information for an energy-loss-vs.-momentum method.
The momenta
can be extracted at the magnetic spectrometer ANKE with high accuracy,
which allows a determination of the excess energy with high precision.
The identification of the $\eta$-meson production itself is done using the
missing mass technique. Here the background subtraction can be performed
using the data obtained at sub-threshold energies. To extract
total and differential cross sections the luminosity is determined
using the simultaneously measured $dp$-elastic scattering.
First data on the excitation function from threshold up to an excess energy
of Q$\sim$10 MeV are shown with an energy resolution of $\Delta$Q$\sim$0.25 MeV.

\newpage

\subsection{  $\eta$-nucleus bound state search at COSY
 }
\addtocontents{toc}{\hspace{2cm}(D. Kirillov)\par}

\vspace{5mm}
D. Kirillov

\vspace{5mm}
\noindent
Institut f\"ur Kernphysik, FZ J\"ulich, Germany
\vspace{5mm}

\noindent
A large acceptance plastic scintillator detector 'ENSTAR'~[1,2] has been designed and built for studies of a new form of nuclear matter - '$\eta$-mesic' nuclei ($\eta$A). These are bound systems, consisting of an  $\eta$-meson and a nucleus. The $\eta$-mesic nuclei, which are solely the result of strong interactions unlike the pionic atoms, are a new kind of atomic nuclei and their research has fundamental significance in studying in-medium properties of hadrons, in particular, medium modification of meson masses. The experimental confirmation of the existence of such $\eta$-bound system will lead to new possibilities of studying the interaction between a nucleus and the short lived (~10-18 s)  $\eta$ meson.

The predicted cross section for such reactions is extremely low (few nbarn) ~[3,4,5]. In order to clearly identify such low cross section events in the presence of a large background from other competitive processes, it is necessary to make coincidence measurements of the ejectile nucleus with the decay products of the $\eta$-mesic nucleus.

The in-beam testing of the detector in full assembled condition was done at COSY, Juelich in March 2004~[6]. Different nuclear reactions (pp elastic scattering, $p + p \rightarrow d + \pi^+$, p + 'heavy target') were used, in addition cosmic ray data were collected. Coincidence data (coincidence between ENSTAR and Big Karl spectrometer, a 2-fold coincidence between different elements in ENSTAR) were also collected.

During the beamtime in May 2005 data on $p + ^{27}Al \rightarrow ^3He + ^{25}Mg_{\eta}$ where obtained. Big Karl was used to spectroscopy and get $\eta$-nucleus missing mass spectra. 'ENSTAR' was used to reduce the background, making  triple coincidences with  $\eta$-mesic nucleus decay products through the chain $\eta + N \rightarrow N^* \rightarrow p + \pi^-$. After the data analysis it was considered, that low statistics does not allow to see the $\eta$-bound nucleus signature. Upper limit for the cross section was preliminary detemined to be 0.7 nbarn.

\newpage

\subsection{Search for the $\eta - {^3\mbox{He}}$ and  $\eta - {^4\mbox{He}}$ bound states 
at COSY-11
 }
\addtocontents{toc}{\hspace{2cm}(Pawel Moskal)\par}

\vspace{5mm}
Pawel Moskal

\vspace{5mm}
\noindent
Institute of Physics, Jagellonian University

\noindent
IKP, Forschungszentrum J\"ulich
\vspace{5mm}

\noindent
Presented talk comprises ideas of the search for the
$\eta - {^3\mbox{He}}$ and  $\eta - {^4\mbox{He}}$ bound states 
by means of the COSY-11 facility at the cooler synchrotron COSY.
Detailes concerning a proposed  method of measurement 
and the preliminary results 
have been reported at the COSY Advisory Committee 
Meeting~\cite{proposal} and at the ETA05 Workshop~\cite{eta05}, and 
the interested reader is referred to these reports. Here instead 
due to the space
limitation only  
abstracts and references are given.

We propose to search for the $\eta - {^4\mbox{He}}$ bound state via a measurement 
of the excitation functions for the $dd \rightarrow {^3\mbox{He}} p \pi^-$ 
and $dd\to {^3\mbox{He}} n \pi^0$ reactions
where the outgoing $p - \pi^-$ or $n -\pi^0$ pair originates
from the conversion of the $\eta$ meson on a neutron inside the ${^4\mbox{He}}$ nucleus
and the ${^3\mbox{He}}$ ejectile is an ``observer''.
Precise determination of the profile of the expected Breit-Wigner distribution 
in the excitation curve will allow  to determine the binding energy and the width 
of the $\eta - {^4\mbox{He}}$ state.  
A simultaneous detection of all ejectiles 
is not feasible in practice neither at the COSY-11  nor at the WASA setup.
However, a kinematically complete measurement can be very well performed
by the detection of  the ${^3\mbox{He}}$ nucleus and one of the other outgoing particle
and  by using a well settled missing mass technique.
For this aim we intend to study the $dd \rightarrow {^3\mbox{He}} p \pi^-$ reaction
at the COSY-11 facility where the outgoing ${^3\mbox{He}}$ and protons can be measured
with a better accuracy  and larger angular range in comparison to the WASA detector.
Besides, under assumption that in the $d-d$ center of mass system the ${^3\mbox{He}}$ 
ejectiles are emitted with the Fermi momenta, the COSY-11 acceptance for detecting ${^3\mbox{He}}$ 
is essentially larger than the one of WASA.
On the other hand, contrary to the COSY-11, the WASA detector allows\ to measure 
the ${^3\mbox{He}} n \pi^0$ decay channel.
Therefore, in the next step in the near future we propose to conduct the measurement of the excitation
function of the $dd\to {^3\mbox{He}} n \pi^0$ reaction
at the WASA-at-COSY facility. 
Both experiments will be performed taking advantage of the possibility
of slow ramping of the COSY deuteron beam.

Preliminary results from $dp \rightarrow {^3\mbox{He}}\,X,\, (X=\pi^0,
\eta$) measurements near the $\eta$ production threshold have been
presented.  The data were taken during a slow ramping of the COSY
internal deuteron beam scattered on a proton target.  The
${^3\mbox{He}}$ ejectiles were registered with the COSY-11 detection
setup.  The ongoing data analysis should deliver high precision data
for the $dp \rightarrow {^3\mbox{He}}\,\eta$ total and differential
cross sections for the excess energies in the range from threshold up
to 9~MeV.  The preliminary excitation function for the reaction $dp
\rightarrow {^3\mbox{He}}\,\pi^0$ does not show any structure which
could originate from the decay of ${^3\mbox{He}}-\eta$ bound state.
We present also a threshold excitation curve for the $dp \rightarrow
{^3\mbox{He}}\,X$ channel.  Contrary to corresponding results from
SATURNE we see no cusp in the vicinity of the $\eta$
threshold~\cite{eta05}.

\newpage

\subsection{On the possibility of measuring the He-$\eta$ bound state with WASA at
COSY
 }
\addtocontents{toc}{\hspace{2cm}(Jozef Zlomanczuk)\par}

\vspace{5mm}
Jozef Zlomanczuk

\vspace{5mm}
\noindent
Department of
Radiation Sciences of Uppsala University,
 Uppsala 27-03-2006
\vspace{5mm}

\noindent
Threshold enhancements in reactions leading to the d-$\eta$
[1,2,3,4], $^{3}$He-$\eta$ [5,6] and $^{4}$He-$\eta$ [7,8] final states
clearly show the low energy $\eta$-nucleus interaction to be strong and
attractive. These enhancements might turn out to be signals for $\eta$-nucleus
quasi-bound or bound states that were predicted for nuclei as light as $^{2}$H
[9] and $^{3}$He [10,11,12,13,14,15]. The effect is expected to be stronger
for larger number of nucleons and some authors expect the $\eta$-nucleus bound
states to exist only for heavier nuclei [16,17,18,19]. Recently, a first
experimental observation of the $^{3}$He-$\eta$ bound state was reported in
the $\eta$ photoproduction reaction on $^{3}$He [20]. Difference between
excitation functions of the $\gamma^{3}$He${}\rightarrow\pi^{0}$p+X reaction
measured for two ranges of the $\pi^{0}$-p relative angle in the $\gamma
$-$^{3}$He center of momentum system, 170$^{0}$-180$^{0}$ and 150$^{0}%
$-170$^{0}$, revealed a structure, which was interpreted as a signature of a
$^{3}$He-$\eta$ bound state with binding energy of a few MeV and width close
to 26 MeV. However, it has been argued that due to limited statistics the
result may be equally well interpreted as a virtual state or merely a cusp
effect [21]. To resolve this ambiguity more accurate measurements are needed.
In order to shed more light on the possibility of existence of $\eta$-nucleus
bound systems for light nuclei, one could study $\eta$-$^{4}$He system in the 
reactions:

dd${}\rightarrow{}^{3}$Hen$\pi^{0}$, dd$\rightarrow{}^{3}$Hep$\pi^{-}$,

\noindent in the energy range from some 60 MeV below dd${}\rightarrow{}^{4}%
$He-$\eta$ threshold to some 40 MeV above threshold. In order to provide small
binning in the total energy, the measurements could be carried out with the
WASA detector on slow ramp of magnetic field of the COSY, resulting in the
beam momentum uniformly distributed over the desired range.

The Monte Carlo simulation of the experiment presented in this contribution
shows that the expected resolution of the WASA detector is sufficient to
unambigously identify the possible $\eta$-$^{4}$He bound state. Also some
estimation of the effect to background ratio is given.

\newpage

\subsection{ Practical unitary $\pi NN$ theory with full dressing}
\addtocontents{toc}{\hspace{2cm}(A. N. Kvinikhidze)\par}

\vspace{5mm}

\noindent
A. N. Kvinikhidze and B. Blankleider.

\vspace{5mm}

\noindent
Mathematical Institute of Georgian Academy of Sciences, Tbilisi,  Georgia,

\noindent
Flinders University of South Australia, Adelaide, Australia.

\vspace{5mm}

\noindent
The complete theoretical solution of the $\pi NN$-like three-body  problem
is a very important theoretical
development which, unfortunately, has so far not been utilized in 
calculations.
The $\pi NN$-like three-body problem is beyond quantum mechanical
considerations as it involves the pure QFT phenomenon of particle  poduction
and
absorbtion. Mathematicaly this is reflected in the presence of extra
disconnected kernels in the $\pi NN$ equations which just correspond  to
pion production
on one nucleon and
absorption on the other, in addition to the purely quanum-mechanical
disconected pair-wise (elastic) interactions. These extra kernels  change
drastically
the dynamics of the three-body system, and in addition, generate  dressings
of the nucleons.
Appart from the difficulty of treating these new kernels to rearrange  the
equations
analogously to the Faddeev rearrangement,  there appears a problem of 
proper renormalization
caused by dressing of the nucleons. The latter cannot be solved using 
traditional
trunction of the Hilbert space to some maximum number of pions.

All these problems are solved, and a formulation of the $\pi NN$ problem
is presented in [1] where unitary equations
are obtained without having to truncate the Hilbert space to some
maximum number of pions. Consequently, all possible dressings of
one-particle propagators and vertices are retained in our model. In
this way we overcome the renormalization problems inherent in
essentially all previous NN theories. The final form of the derived 
equations
is very convenient for numerical solution as basically the same  methods as
applied to the Faddeeev equations are needed.

\newpage

\subsection{ Full two body amplitudes as input for 
the $\eta$ meson production in nucleon-nucleon scattering }
\addtocontents{toc}{\hspace{2cm}(A. \v{S}varc)\par}

\vspace{5mm}

\noindent
A. \v{S}varc, S. Ceci and B. Zauner

\vspace{5mm}

\noindent
Ru\dd er Bo\v skovi\' c Institute, Bijeni\v cka cesta 54, 10000 Zagreb, Croatia.

\vspace{5mm}

\noindent
Theoretical models for calculating meson production processes in
nucleon-nucleon scattering require two-body  meson-nucleon
amplitudes with at least one particle off-mass-shell as input. As
experiments enable us only to obtain  the amplitudes with all four
particles on-mass shell, we need the model for the off-mass-shell
extrapolation. In  principle, one should discuss the full invariant
amplitude, in practice it is done on the level of partial waves 
only. Early attempts have been done in $p  p \rightarrow \pi d$ process
\cite{Gre84,Bat94} where the on-shell amplitudes with the $\pi N$ cm
energy calculated with the full off-mass-shell kinematics were used as
the off-mass-shell values. These  attempts have been followed by
representing the S$_{11}$ partial wave within the framework of
separable potential  model. In this approach the free
model-parameters are obtained by fitting PWA results, and are used to
calculate the  input meson-nucleon form factors using the full
off-mass-shell kinematics \cite{Gar02,Gar05}.

We offer the explicit analytic form of the partial-wave T-matrix emerging from the two-body coupled-channel formalism 
\cite{Bat98,Sva06}, which  describes all available experimental data in $\pi N$ elastic and \piNetaN channels, and 
may therefore serve as a basic expression which determines the behavior of partial wave amplitudes when one or two 
particles are going off-mass-shell:
\be
      \hat{T}(s, q_i , q_f) =   \sqrt{Im  \hat{\Phi}(q_i)} \cdot
    \hat{\gamma}^{\rm T} \cdot  \frac{\hat{G}_{0}(s)}{I- \left[\hat{\gamma} \cdot \hat{\Phi}(q_i,q_f) \cdot 
\hat{\gamma}{\rm ^T}\right] \cdot \hat{G}_{0}(s)} \cdot  \hat{\gamma}  \cdot \sqrt{Im  \hat{\Phi} (q_f) }.
\label{eq:final} \nonumber
      \ee
	$q_i$ and $q_f$ are the incoming and outgoing meson-nucleon cm momenta and s is the meson-nucleon cm energy.  
This is a symbolic matrix equation with free parameters contained in the channel-resonance coupling matrix 
$\hat{\gamma}$ and in the values of the free Green function propagator poles.  The recommended representation of the 
on-shell experimental data in principle fully defines the recipe how to go off-mass-shell with either particle. 

This recipe, however, requires to be tested; so we encourage all physicists involved in any theoretical 
considerations involving $\eta$-meson production to use as input to their calculation. In order to simplify your 
efforts,
Mathematica 5.2 codes for obtaining the analytic form of partial wave T-matrices, together with the required input 
parameter file, or the Mathematica 5.2 table $T_{ab}(i,j,k)$ for the $\hat{T}(s, q_i , q_f)$ matrix with required 
density are available upon request for all partial waves \cite{Sva06a}, and will be send to the interested person 
immediately.

\newpage

\subsection{ Why the $\eta$-nucleus scattering length can be small?
 }
\addtocontents{toc}{\hspace{2cm}( J. A. Niskanen)\par}

\vspace{5mm} J. A. Niskanen

\vspace{5mm}
\noindent
Department of Physical Sciences, PO Box 64, FIN-00014
University of Helsinki, Finland
\vspace{5mm}

\noindent
Most calculations of $\eta^3$He scattering yield rather large
imaginary parts of the scattering length, mostly $\Im a\geq 2$ fm
and $\Im a\geq |\Re a|$ [1--4]. Even if $\Re a$ were negative,
considered often as a sign of possible binding, this is bad news for
finding bound states. Firstly, the bound state could be too broad to
observe and secondly even the actual necessary conditions for the
existence of a bound state $|\Re a| > \Im a$ [5] (expanded further
to $\Re [a^3(a^*-r_0^*)] > 0$ in Ref. [6]) are not satisfied.

However, two recent global analyses of $pd\to\eta^3$He data give
mutually consistent results $a = \pm 4.3 \pm 0.3 + i\, (0.5 \pm
0.5)$ fm [6] and $a= 4.24\pm 0.29 + i\, 0.72\pm 0.81$ fm [7] with
intriguingly small imaginary parts. This is somewhat surprising,
since ad hoc one might naively expect absorption on three nucleons
to be three times as much as on a single nucleon and $\Im a\geq 1$
fm. Further, in addition to elementary absorption $\eta N
\rightarrow \pi N$ there are new channels, notably absorption on two
or three nucleons.

Ref. [8] considers several of these contributions. Firstly, it is
pointed out that the presumably dominant part of quasifree
elementary absorption (assumed to be coherent from a bound state
extending all over the nucleus) should be proportional to the total
isospin operator on the nucleons. Therefore, in the case of $^3$He
this would give essentially the same result as a single nucleon, not
three times as much. Consideration of nuclear inelasticities on two
or three  nucleons are seen to be also small, even negligible in
comparison with quasifree absorption. Absorption where the nucleus
remains intact $\eta^3{\rm He} \rightarrow \pi^+t$ was seen to be
small already in Ref. [6].

Consequently, there does not seem to be any pressing need for $\Im
a$ much larger than in the elementary $\eta N$ case in agreement
with [6] and [7]. Similar arguments may be extended to other nuclei,
too, probably improving the prospects of searches for bound states.

\newpage


\subsection{
 Issues in $np\to\eta d$ near threshold}
\addtocontents{toc}{\hspace{2cm}(H. Garcilazo)\par}

\vspace{5mm}

H. Garcilazo $^{(1)}$
and M. T. Pe\~na $^{(2)}$

\vspace{5mm}

\noindent
$(1)$ 
Instituto Polit\'ecnico Nacional, Edificio 9,
07738 M\'exico D.F., Mexico

\noindent
$(2)$  Instituto Superior T\'ecnico,
Av. Rovisco Pais, P-1049-001 Lisboa, Portugal

\vspace{5mm}

\noindent
The $np\to\eta d$ cross section near threshold was measured in
Refs. [1,2].
The model to describe this process has been described in detail
in Refs. [3-7]. Our main conclusion was that in order to explain the 
energy dependence of the $np\to\eta d$ cross section near threshold the
real part of the $\eta N$ scattering length must be between 0.42 and
0.72 fm. In this talk we will concentrate on three issues:
\noindent
i) Effect of the $\sigma$ width. Our three-body model is based in the 
meson-nucleon two-body channels $\eta N$-$\pi N$-$\sigma N$ where the 
$\sigma N$ channel represents the inelastic $\pi\pi N$ channel and it
has been included as an effective channel since the $\sigma$ was 
taken to be a stable particle of mass $m_\sigma=2m_\pi$. We have now
shown that taking the $\sigma$ as a $\pi\pi$ resonance with mass 
and width determined by the $I=J=0$ phase shift [8] does not change
qualitatively the results. 
\noindent
ii) Can one pin-down the $\eta N$ scattering length? The seven models 
of the $\eta N$ amplitude that we use have been obtained in Refs.
[9-12]. They are characterized by having 
$Re(a_{\eta N})$=0.42, 0.72, 0.75, 0.83, 0.87, 1.05, and 1.07 fm where
$a_{\eta N}$ is the $\eta N$ scattering length which together with the
effective range $r_{\eta N}$ and shape parameter $s_{\eta N}$
describe the $\eta N$ scattering amplitude at low energies. 
We consider $Re(a_{\eta N})=x$ and take $Im(a_{\eta N})$, $Re(r_{\eta N})$,
$Im(r_{\eta N})$, $Re(s_{\eta N})$, and $Im(s_{\eta N})$ as functions
if $x$ which allows us to generate models of the $\eta N$ amplitude for
arbitrary values of $x$. Repeating our analysis with these fictitous
models we conclude that $0.47 \le Re(a_{\eta N})\le 0.64$ fm.
\noindent
iii) Quasivirtual versus quasibound $\eta NN$ state. In order to
clarify the nature of the pole near threshold present in the $\eta NN$ 
system we calculated the $\eta d$ elastic scattering amplitude [7] and
obtained from it the effective-range parameters $a_{\eta d}$, $r_{\eta d}$,
and $s_{\eta d}$. Using this effective-range expansion we then located
the nearest pole in the complex $q$-plane and found in all cases that 
the pole is located in the third quadrant which means that it is a
quasivirtual state.

\newpage

\subsection{ Chiral approach to $\eta$ in a nuclear medium
 }
\addtocontents{toc}{\hspace{2cm}(E. Oset)\par}

\vspace{5mm}
E. Oset

\vspace{5mm}
\noindent
Departamento de F\'\i sica Te\'orica and IFIC, Universidad de Valencia
\vspace{5mm}

\noindent
Basic elements of chiral dynamics implementing unitarity in coupled channels
were presented, in particular for the case of the interaction of the
octet of
pseudoscalar mesons and the octet of stable baryons, which leads to a set of
dynamically generated resonances, like the $\Lambda(1405)$,
$\Lambda(1670)$,
$N^*(1535)$, etc. \cite{Oset:1997it,Oset:2001cn,Inoue:2001ip}.
  The $N^*(1535)$ resonance appears with a width of about 95 MeV, like
proposed
at BES \cite{Bai:2001ua}, which is narrower than in other approaches.  The
discussion focused on whether the models that lead to large widths of the
$N^*(1535)$ by making fits to the data could not give results equally
acceptable
by making constraint fits with the $N^*(1535)$ width given by the
precise BES
experiment.

    With this chiral model where the  $N^*(1535)$ is dynamically
generated and
one gets reasonable results for the $\eta N$ interaction, one constructs an
$\eta$ nucleus optical potential which has a manifest energy dependence
\cite{Inoue:2002xw}, qualitatively similar to another one obtained on more
phenomenological grounds \cite{Post:2003hu}, and with this potential one
solves the Klein Gordon
equation searching for bound $\eta$ states in different nuclei. This is done
in \cite{Garcia-Recio:2002cu} with the result that there are indeed bound
states in nuclei heavier than $^{12}C$, but probably also lighter where
it was not
checked, but the widths are inevitably larger than the energy separation
between
the levels, what should make the identification of these states a non
trivial
task.

  To finish the discussion a call was made to the paper of Nagahiro,
Jido and
Hirenzaki \cite{Nagahiro:2003iv} where the production of $\eta$ bound
states is
studied theoretically by means of the recoilless $(d,^3He)$ reaction.  It is
found there that the spectra does not show any visible sign of the bound
states
because they are so broad, and on the other hand a peak structure with a
width
of about 30 MeV is generated because of the recoilless $(d,^3He)$
reaction, which
magnifies the cross section where the recoil momentum is about zero, and
makes
it decrease to lower and higher energies as one diverts from the magic
recoilless
kinematics.  This should serve as a warning that not every peak seen in an
experiment looking for bound $\eta$ has to correspond to a bound state.
On the
other hand this paper also shows that the shapes of the cross section
depend on
the size of the optical potential, so, hopes are given that indirectly,
but not
through the peak structure, one might come to learn something about the
interaction of $\eta$ with nuclei.

\newpage

\subsection{$\eta$ Photoproduction on the Nucleus\protect\footnote{Work
supported by DFG}}
\addtocontents{toc}{\hspace{2cm}(U. Mosel)\par}

\vspace{5mm} U. Mosel\footnote{mosel@physik.uni-giessen.de}

\vspace{5mm}
\noindent
Institut f\"ur Theoretische Physik, Universt\"at Giessen, Giessen, Germany
\vspace{5mm}

\noindent
Eta photoproduction on heavier nuclei has been studied in a model
that takes both the primary interaction and the final state
interactions (FSI) into account. This model contains 2 steps. In the
first step the initial interaction of the incoming photon with any
of the nucleons of the target nucleus is treated by using cross
sections for photons interacting with free nucleons. In this step
'trivial' in-medium effects are taken into account, such as Fermi
motion of the hit nucleon and Pauli-blocking. The mesons produced in
this interaction are then propagated through the nuclear medium
until they leave the nucleus.  Essential for this step is that it
allows not only for absorption of produced mesons, but also for
elastic and inelastic rescattering. In the latter process
channel-coupling effects are taken fully into account so that, for
example, the finally observed eta has not necessarily been produced
in a first interaction step, but can have been produced later
through an inelastic rescattering process of an originally produced
pion.

In this latter step of final state propagation in-medium
selfenergies can be included and their effect can be studied.
Selfenergies of the eta and the relevant nucleon resonances have
been calculated in \cite{Lehr}. There it has been found that the
in-medium changes of the selfenergy of the N(1535) are relatively
minor, the biggest contribution coming from the small $\rho$ decay
branch of this resonance that opens up when the $\rho$ meson becomes
softer in medium. In a selfconsistent coupling model \cite{Post}
that involves various resonances and mesons relevant for this energy
regime we find that $\eta$ mesons in nuclei should experience an
attractive potential of about -40 MeV that opening the possibility
of$\eta$ bound states in nuclei.

The final state interactions have been modeled using the GiBUU
semiclassical transport model \cite{GiBUU}. The model has been
compared with the TAPS data on nuclei \cite{Hombach,Lehr} and gives
in general a good description of the experimental results as far as
the total cross sections go. The data taken at higher energies
exhibit a clear sensitivity to the in-medium potential of the
N(1535) and in particular its momentum dependence. However, the
energy-differential cross sections $d\sigma/dT_\eta$ still present a
puzzle. Here the calculated distributions are shifted significantly
to higher energies, compared to experiment \cite{Effenberger}. This
is true, when the FSI cross sections are taken consistently from a
resonance model. Assuming instead a constant FSI cross section of
$\sigma^\eta_{\rm inel} = 30$ mb does describe the data. An
explanation for this problem is so far not available.

The importance of treating coupled channel effects correctly in the
FSI becomes clearly visible in calculations for $\eta$
electroproduction with virtual photons. Here our calculations show
that while at the photon point secondary production processes are
negligible, at higher virtualities of the incoming photon the
process $\gamma + N \to \pi + N$, $\pi + N' \to \eta + N'$ becomes
dominant \cite{Lehrelectro,Lehr1}.

\newpage

\subsection{ 
Eta-meson production on the nucleon within the Giessen model
\protect\footnote{Supported
by Forschungszentrum Juelich}}
\addtocontents{toc}{\hspace{2cm}(V. Shklyar)\par}

\vspace{5mm}

V. Shklyar, H. Lenske, and U. Mosel

\vspace{5mm}

\noindent
Institut f\"ur Theoretische Physik, Universit\"at Giessen, D-35392
Giessen, Germany

\vspace{5mm}

\noindent
In the Giessen model pion- and photon-induced reactions in the
nucleon resonance energy region are described by an unitary
coupled-channel effective Lagrangian approach
[1-6]
The coupled-channel problem is treated in the Bethe-Salpeter
formalism where the interaction kernel is constructed from the
effective interaction Lagrangians (see [6] and
references therein). The $K$-matrix approximation is used  to solve
the Bethe-Salpeter equation maintaining unitarity as well as Lorenz
and gauge invariance.

The  resonance part of the interaction kernel relevant for the $\eta$-production consists of
$S_{11}(1535)$, $S_{11}(1650)$, $P_{11}(1440)$, $P_{11}(1710)$, $P_{13}(1720)$, $P_{13}(1900)$,
 $D_{13}(1520)$, $D_{13}(1850)$, $D_{15}(1676)$, $F_{15}(1680)$, $F_{15}(2000)$ resonances.
The resonance and background parameters are constrained by
experimental data from a number of reactions: $\pi N\to \pi N$,
$2\pi N$, $\eta N$, $\omega N$, $K\Lambda$, $K\Sigma$ and $\pi N\to
\gamma N$, $\pi N$, $\eta N$, $\omega N$, $K\Lambda$, $K\Sigma$. The
main difference to our previous calculations is the inclusion of
higher partial waves with spin-$\frac{5}{2}$ which were omitted in
[1-4]. The
updated $\pi N \to\eta N$ and $\gamma p \to \eta p$ transition
amplitudes are obtained by solving the multichannel problem as
explained in [6-7]. Although the
spin-$\frac{5}{2}$ resonances are found to be coupled only weakly to
the final $\eta N$ final state [5], these
contributions can be important for the description of the photon beam
asymmetry because of interference effects.

Similar to our previous findings [3] the  $\pi^- p
\to \eta n$ reaction at c.m. energies 1.48...1.65 GeV is dominated
by the $S_{11}$-partial wave contributions. Overall, the
$S_{11}(1535)$-state plays the dominant role but above 1.65 GeV the
destructive interference with the second $S_{11}(1650)$-resonance
decreases the effective contribution from this partial wave. A
similar behaviour was also found in the calculations of J\"ulich
group [8]. In the present study the
$P_{11}(1710)$-resonance is found to be completely inelastic. This
state together with the background contributions dominate the $\pi^-
p \to \eta n$ reaction at energies 1.6 .. 1.8 GeV.

A much larger database is available for the $\gamma p\to\eta p$
reaction [9-14].
In the energy region up to 1.75 GeV the $\eta$-photoproduction cross
section on the proton is entirely dominated by the $S_{11}(1535)$
resonance contribution. At higher energies the photoproduction cross
section is a sum of different partial waves without a single
dominant partial wave. This conclusion is in line with findings of
[15,16].

One of the major questions to be addressed in future studies are the
magnitudes of the proton and neutron helicity amplitudes of the
$S_{11}(1535)$-resonance. While MAID and Eta-MAID conclude on the
same ratio of $A^n_{1/2}/{A^p_{1/2}}\approx$ 0.8, the absolute value
of the  proton helicity amplitude is different in these studies. At
the same time the GW-analysis of the pion photoproduction data
[17] gives a smaller value for
$A^n_{1/2}/{A^p_{1/2}}\approx$ 0.33. We consider this as strong
support for a combined analysis of pion- and eta-photoproduction in
order to understand the property of this resonance. This study
becomes even more interesting in view of forthcoming data on
quasi-free eta-photoproduction on a neutron in a deuterium target
planned by CB-ELSA and GRAAL collaborations.

\newpage

\subsection{ 
Model dependence of the low-energy eta-nuclear interaction
 }
\addtocontents{toc}{\hspace{2cm}(A. Fix)\par}

\vspace{5mm}
A. Fix and H. Arenh\"ovel

\vspace{5mm}
\noindent
Institut f\"ur Kernphysik, Johannes Gutenberg-Universt\"at, Mainz, 
Germany
\vspace{5mm}

\noindent
At present, our main knowledge about the low-energy $\eta N$ interaction
is obtained from the analysis of $\pi N\to \pi N$, $\pi N\to \eta N$
and other reactions. A typical calculation is based on a coupled
channel $K$- or $T$-matrix approach [1,2,3].
However, currently available models still provide quite weak
constraints of the $\eta N$ amplitude. 
An alternative method to study the $\eta N$ interaction is to
extract the corresponding information from $\eta$ production
data. Due to the short range nature of the main mechanisms of such
processes the energy dependence of their cross sections is mainly
governed by the long range part of the final state wave functions.
This feature makes it possible to extract reliable information 
on low-energy $\eta$-nuclear scattering.

As an adequate way to embed the $\eta N$ dynamics into a microscopic
$\eta$-nucleus calculation we solve the few-body scattering equations. 
This task is facilitated by using separable representations of the 
of the integral kernels. The approach allows one to
reduce the dynamical equations to a numerically manageable form
without significant loss of the basic physics. 
The reduction scheme is presented
in [4,5,6] for the three-body system $\eta NN$ and in [7]
for the $\eta$-$3N$ case.

The main question we have focused on is 'How sensitive are the
few-body results to the details of the elementary $\eta N$
interaction?' This concerns primarily the part of the $\eta N$ phenomenology 
which are not well controlled by the $\eta N$ analyses.
In order to understand the manner in which the properties of $\eta N$
interaction can influence $\eta$-nuclear phenomena the 
following issues were addressed:
\begin{itemize}
\item Sensitivity of the $\eta d$ scattering length $a_{\eta d}$ to
the variation of the $\eta N$ amplitude $t_{\eta N}(\vec{p},\vec{p}\,')$
at short distances in the $\eta N$ system.
\item Dependence of $a_{\eta d}$ on the model for the
$S_{11}(1535)$ resonance.  
\end{itemize}

With respect to the first point, 
the matrix $t_{\eta N}(\vec{p},\vec{p}\,')$ was
modified through multiplication by the step functions
$\theta(p-p_c)$ and $\theta(p'-p_c)$, so that $t_{\eta N}$ was set
to zero if at least one of its momentum arguments exceeds the cut-off
value $p_c$. We find a strong sensitivity of $a_{\eta d}$ for $p_c <
0.5$ fm$^{-1}$ and essentially no dependence on $p_c$ in the region
$p_c > 1$ fm$^{-1}$.

For the second point, we compared the results for $a_{\eta d}$ obtained for
three different models of the $\eta N$ interaction: (i) an isobar model in
which $S_{11}$ is mainly treated as a genuine baryon resonance
[3], (ii) the potential model from [2], where the $\eta N$
resonance in the $s$-wave is generated through an artificial barrier
in the $\eta N$ potential, and (iii) a dynamical model, where 
the $S_{11}(1535)$ is a quasibound $K\Sigma$ state appearing as a resonance in 
the $\eta N$ channel [8,9]. In each case the model
parameters were fitted to the same value of the $\eta N$ scattering
length and the same physical mass of the $S_{11}(1535)$ resonance.
Our calculation shows that all three models lead to quite similar
values of $a_{\eta d}$. 
As next step, we studied the sensitivity of the $\eta d$
scattering length to the $\eta N$ amplitude in different regions of
the $\eta N$ energy $E$. Our results showed
that only a rather small region $-20
< E < 0$ MeV, where the $\eta N$ amplitude $f_{\eta N}$ is close to its zero 
energy value $a_{\eta N}$, is responsible for the resulting value of 
$a_{\eta d}$.

In view of the similarity of the $\eta d$ scattering length $a_{\eta
d}$ arising from different $\eta N$ approaches and the irrelevance of
the short range $\eta N$ dynamics we conclude that low-energy $\eta
$-deuteron (and probably $\eta\,^3$He) properties are sensitive to
the behavior of the $\eta N$ off-shell $t$-matrix only over a
limited range of its momentum and energy arguments. This conclusion tells us
that studies of $\eta NN$ and $\eta$-$3N$ phenomena will
provide reliable information about the low-energy parameters of the
$\eta N$ interaction.

\newpage

\subsection{The role of dynamically generated resonances in $\boldmath{\eta}$ production
 }
\addtocontents{toc}{\hspace{2cm}(M. D\"oring)\par}

\vspace{5mm}
M. D\"oring\footnote{Email: doering@ific.uv.es}, E. Oset, and D. Strottman 

\vspace{5mm}
\noindent
IFIC, University of Valencia, Spain

\vspace{5mm}

\noindent
The unitary extensions of chiral perturbation theory $U\chi PT$ have 
brought new
light in the study of the meson-baryon interaction and have shown 
that some well
known resonances qualify as dynamically generated, or in other 
words, they are
quasibound states of a meson and a baryon, the properties of which 
are described in
terms of chiral Lagrangians. After early studies in this direction 
explaining the
$\Lambda (1405)$ and the $N^*(1535)$ as dynamically generated resonances
\cite{Kaiser:1995cy,Kaiser:1996js,kaon,Nacher:1999vg,Oller:2000fj}, 
more systematic
studies have shown that there are two octets and one singlet of 
resonances from the
interaction of the octet of pseudoscalar mesons with the octet of 
stable baryons
\cite{Jido:2003cb,Garcia-Recio:2003ks}. The $N^*(1535)$ belongs to 
one of these two
octets and plays an important role in the $\pi N$ interaction with its coupled
channels $ \eta N$, $K \Lambda$ and $K \Sigma$ \cite{Inoue:2001ip2}. 
In spite of the
success of the chiral unitary approach in dealing with the 
meson-baryon interaction
in these channels, the fact that the quantum numbers of the $N^*(1535)$ are
compatible with a standard three constituent quark structure and that 
its mass is
roughly obtained in many standard quark models 
\cite{Isgur:1978xj,Capstick:1993kb}, or
recent lattice gauge calculations \cite{Chiu:2005zc}, has as a 
consequence that the
case for the $N^*(1535)$ to be described as a dynamically generated 
resonance appears
less clean than that of the $\Lambda (1405)$ where both quark models 
and lattice
calculations have shown systematic 
difficulties\cite{Nakajima:2001js}. 

\vspace*{0.2cm}

In the model of \cite{Inoue:2001ip2} the $N^*(1535)$ appears as a dynamically generated resonance after fitting the subtraction constants of the intermediate meson-baryon loops to $\pi N$ data in the channels $S_{11}$ and $S_{31}$. It is remarkable that through the inclusion of the $\pi\pi N$ channel also the $\Delta(1620)$ appears as a pole in the complex plane of $\sqrt{s}$ in the $S_{31}$ channel. The model gives also a good description of the $\pi N\to\eta N$ cross section near the $\eta N$ threshold (further above, the $N^*(1535)$ is a bit too narrow). This is a consequence of unitarity and the inclusion of the $\pi\pi N$ channel in the model: the latter channel is the only open inelastic channel relevant in the energy region of the $N^*(1535)$; as the $\pi\pi N$ channel is well described by the model, the reproduction of the $\pi N\to\eta N$ data is automatic. Thus, further information is desirable to isolate the molecular components of the $N^*(1535)$. In this context a calculation of the electro-production of the $N^*(1535)$ would offer an interesting and independent test of the model of \cite{Inoue:2001ip2}.

\vspace*{0.2cm}

In Ref. \cite{Baru:2003qq} it has been pointed out that the scattering length and effective range of the $\bar{K}K$ interaction give hints of the size of the molecular component of the $f_0(980)$. In particular, a positive effective range is a sign of a molecular state. 
For the $N^*(1535)$ the situation is different as in the unitary chiral model \cite{Inoue:2001ip2} this resonance appears as a quasibound state of the $K\Sigma$ and $K\Lambda$. However, for further studies the threshold parameters have been provided in the workshop and are displayed in Tab.
\ref{tab:effrange} and \ref{tab:effrange2}. For a recent compilation on the $\eta N$ scattering length see Ref. 
\cite{Sibirtsev:2001hz}.
\begin{table}
\begin{tabular}{lllll}
\hline\hline Channel&Re $a$ [fm]&Im $a$ [fm]&Re $r_0$ [fm]&Im $r_0$ [fm]
\\
\hline
$K^+\Sigma^-\to K^+\Sigma^-$&$-0.29$&$+0.087$&$+0.58$&$-1.50$
\\
$K^0\Sigma^0\to K^0\Sigma^0$&$-0.21$&$+0.067$&$-2.25$&$-0.36$
\\
$K^0\Lambda\to K^0\Lambda$&$-0.15$&$+0.17$&$+0.74$&$-3.21$
\\
$\pi^-p\to\pi^-p$&$+0.080$&$+0.003$&$-14.7$&$-22.3$
\\
$\pi^0n\to\pi^0n$&$-0.023$&$0$&$-31.4$&$0$
\\
$\eta n\to\eta n$&$+0.27$&$+0.24$&$-7.26$&$+6.59$
\\
\hline\hline
\end{tabular}
\caption{Scattering lengths $a$ and effective range parameters $r_0$ calculated from the model \cite{Inoue:2001ip2}.}
\label{tab:effrange}
\end{table}

\begin{table}
\begin{tabular}{llll}
\hline\hline Channel&$a$ [fm]&$a_{\rm exp.}$ [fm]&$a$ [fm]
\\
\hline
{\small$ K\Sigma\to K\Sigma$, $I=1/2$}&$-0.12+i\,0.03$&&$-0.15+i\,0.09^{(*)}$
\\
{\small$K\Sigma\to K\Sigma$, $I=3/2$}&$-0.34+i\,0.10$&&$-0.13+i\,0.04^{(*)}$
\\
{\small$K\Lambda\to K\Lambda$}&$-0.15+i\,0.17$&&$+0.26+i\,0.10^{(*)}$
\\
{\small$\pi^-p\to\pi^-p$}&$+0.080+i\,0.003$&$+0.123\pm 0.02^{(**)}$&
\\
{\small$\pi^0n\to\pi^0n$}&$-0.023+i\,0$&$-0.004\pm 0.006^{(***)}$&
\\
{\small$\eta n\to\eta n$}&$+0.27+i\,0.24$&&{\footnotesize $\begin{array}{l}+0.43+i\,0.21^{(*)}\\ +0.717+i\,0.265^{(****)}\\
+0.20+i\,0.26^{(*****)}\end{array}$}
\\
\hline\hline
\end{tabular}
\caption{Scattering lengths $a$ from unitary coupled channel approach \cite{Inoue:2001ip2} (second column), ''experiment'' (third column), other work (last column).
\newline
{\footnotesize
$(*)$ Lutz, Wolf,Friman, NPA {\bf 706}
\newline
$(**)$ Ericson, Loiseau, Wyech, hep-ph/0310134
\newline
$(***)$ M.D\"oring., Oset, Vicente Vacas, PRC {\bf 70}
\newline
$(****)$ A. Svarc
\newline
$(*****)$ Kaiser, Waas, Weise, {\bf NPA 612}
}
}
\label{tab:effrange2}
\end{table}

\vspace*{0.2cm}

In Refs. \cite{Doring:2005bx} and \cite{Doring:2006pt} the $\eta$
production has been investigated for a larger set of pion- and
photon-induced reactions. For the $\gamma p\to\pi^0\eta p$ reaction
the $N^*(1535)$ has again been found important, although cross section
and invariant mass spectra are almost independent on the actual width
of the resonance as it becomes clear when using a phenomenological
$\pi N\to\eta N$ $s$-wave transition potential that exhibits a larger
width.

At the energies of the $\gamma p\to\pi^0\eta p$ reaction other
resonances, which couple to the $\eta$, certainly play an important
role.  The $\Delta^*(1700)$ resonance qualifies as dynamically
generated through the interaction of the $0^-$ meson octet and the
$3/2^+$ baryon decuplet as recent studies show
\cite{Kolomeitsev:2003kt,Sarkar:2004jh}. In this picture it is
possible \cite{Sarkar:2004jh} to obtain the coupling of the
$\Delta^*(1700)$ to the $\eta\Delta(1232)$ and $ K \Sigma^*(1385)$ for
which experimental information does not yet exist.  For the $\gamma
p\to\pi^0\eta p$ reaction this resonance turns out to play an
important role. Comparing the predictions with preliminary data from
\cite{nanovanstar}, a good agreement is found. The $\Delta^*(1700)$
has been studied in \cite{Doring:2006pt} for the reactions $\pi^- p\to
K^0\pi^0\Lambda$, $\pi^+ p\to K^+\pi^+\Lambda$, $K^+\bar{K}^0p$,
$K^+\pi^+\Sigma^0$, $K^+\pi^0\Sigma^+$, and \underline{$\eta\pi^+ p$},
in which the basic dynamics is again given by the excitation of the
$\Delta^*(1700)$ resonance which subsequently decays into
$K\Sigma^{*}(1385)$ or $\Delta(1232)\eta$.  In a similar way the
$\gamma p\to K^0\pi^+\Lambda$, $K^+\pi^-\Sigma^+$, $K^+\pi^+\Sigma^-$,
and $K^0\pi^0\Sigma^+$ related reactions are studied. Besides a good
description of the energy dependence of the cross sections in most of
the reactions, for all reactions a good global agreement is found at
energies up to around $s^{1/2}=1930$ MeV which is remarkable as the
various cross sections differ by a factor of 50 for the pion-induced
reactions and by a factor of 5-10 for the photon-induced ones. Thus,
the data of the various three-body final states, among them $\eta\pi
N$, gives support to the $\Delta^*(1700)$ being a dynamically
generated resonance.

\newpage

\subsection{$\pi N \to \eta N$ reaction within the framework of the coupled-channels meson-nucleon model
 }
\addtocontents{toc}{\hspace{2cm}(A. Gasparyan)\par}

\vspace{5mm}
{A. Gasparyan$^{1*}$, J. Haidenbauer$^2$, and C. Hanhart$^2$}

\vspace{5mm}
\noindent
$^1$Institute of Theoretical and Experimental Physics,
Moscow, Russia

\noindent
$^2$Institut f\"ur Kernphysik (Theorie), Forschungszentrum J\"ulich,
J\"ulich, Germany

\vspace{5mm}

\noindent
A  coupled-channels meson-exchange model for the elastic and inelastic $\pi N$ scattering
developed by the J\"ulich group is presented ~[1]. The model includes
channels $\pi N$, $\eta N$, as well as three effective
$\pi\pi N$ channels namely  $\rho N$, $\pi\Delta$, and $\sigma N$.
The model is based on the effective potential, containing $t$-,$u$-, and
$s$- channel exchange diagrams. This potential is used then in the Lippmann-Schwinger
equation in order to obtain the scattering amplitude. The goal is to describe the $\pi N$
data not only close to $\pi N$ threshold but also in the resonance region.

There exist many models describing  $\pi N$ scattering in the resonance
region. Those include simple resonance parameterization~[2],
$K$-matrix approximation~[3], separable models~[4], and chiral coupled-channel
approach~[5,6,7].
Compared to most of them, the advantages of the J\"ulich model
are that the background is fitted to all partial waves simultaneously and that
resonances can be generated dynamically (in particular, the Roper resonance is 
generated due to  a strong interaction in the $\sigma N$ channel) or they can be included
as bare s-channel poles. 

The resulting phase shifts and inelasticities for isospin $1/2$ and $3/2$ and
$j\le 5/2$ partial waves describe the empirical data~[8] rather well up to
the c.m. energy $1900$ MeV. The total $\pi N\to \eta N$
cross section~[9-13] is also well reproduced. It turned out that the $\pi\pi N$ channel
is necessarily needed if one wants to simultaneously describe the $\pi N\to \eta N$
cross section and the $S_{11}$ inelasticity in the $\pi N$ channel.
One gets for the $\eta N$  scattering length (which is a very important
quantity in the $\eta$-nucleus physics) the value $a_{\eta N}=0.4+i0.25$ fm.
The $\pi N\to \eta N$ differential cross section~[9,12,13] is reproduced up to
the energies around $1650$ MeV. At higher energies the agreement with
the experimental data is worse. A possible explanation for that is
the need to include higher partial waves coupling to the $\eta N$
channel.

For the consistency of the model the $\omega N$ channel and
strangeness channels also have to be included. The work along this line is in progress now.

\newpage

\subsection{Three-body model calculations for the final state of
the  $np\to\eta d$ reaction near threshold
 }
\addtocontents{toc}{\hspace{2cm}(M. T. Pe\~na)\par}

\vspace{5mm}
M. T. Pe\~na $\,^{(1),*}$ and H. Garcilazo $\,^{(2)}$

\vspace{5mm}

\noindent
$(1)$  Instituto Superior T\'ecnico,
Av. Rovisco Pais, P-1049-001 Lisboa, Portugal

\noindent
$(2)$ 
Instituto Polit\'ecnico Nacional, Edificio 9,
07738 M\'exico D.F., Mexico
\vspace{5mm}

\noindent
We present in this talk the results of three-body model calculations [1,2] for the
$np\to\eta d$ reaction near threshold.
Since the existing empirical knowledge of the $\eta$N interaction and
S11-N$^*$(1535) resonance provides a dispersion
of the $\eta N$ scattering 
length in the interval $0.27 \rm fm \le Re(a_{\eta N}) \le 1.05 \rm fm$,
our calculations probed different $\eta N$ dynamical models,
based upon recent data analysis of the coupled reactions 
$\pi N {\to} \eta N$, $\eta N {\to} \eta N$ and
$\gamma  N {\to} \eta N$ [3-6].

For each model, the short-range production mechanism strength was fitted to
the very-near-threshold
region, and from there the cross section
at higher energies was predicted. Only the J\"ulich model [3] 
describes the data reasonably well throughout the full 
energy range. The preference of the data for
a model with small Re$(a_{\eta N})$, as the J\"ulich model, is to a large extent 
independent of the production mechanism, and our results show  that the
characteristic shape  near threshold of the experimental 
cross section is a signature
of the $\eta d$ final-state interaction. On the contrary, the absolute value 
of the cross section depends on the production  mechanism.

In this presentation
we also report how these results
are modified by a relativistic treatment
for the kinematics of the 3 particles involved, and the boosts
of the two-body interactions.
In common, both the relativistic and non-relativistic calculations
show that the shape of the cross-section is essentially determined
by the $\eta$d three-body final state interaction alone.
The relativistic calculation indicates that the description of the
$np{\to}\eta d$ reaction near threshold implies that the $\eta N$ scattering
length is in the interval $0.42<a_{\eta N}<0.72 \rm fm$. In the following lecture by
Humberto Garcilazo, it is shown how to narrow this uncertainty interval.

\newpage

\subsection{An Isobar Model for $\eta$ Photo- and Electroproduction on the Nucleon
 }
\addtocontents{toc}{\hspace{2cm}(L. Tiator)\par}

\vspace{5mm}
L. Tiator

\vspace{5mm}
Institut f\"ur Kernphysik, Universt\"at Mainz, Mainz, Germany
\vspace{5mm}

\noindent
ETAMAID is an isobar model for eta photo- and electroproduction on the nucleon. It is
available as an online program on the web~[1] for easy access and individual studies of
model parameters and kinematical conditions.

The isobar model contains a field-theoreti\-cal background with $s$- and $u$-channel
nucleon Born terms and $\rho,\omega$-exchange in the $t$-channel. In the resonance sector
eight nucleon resonances in Breit-Wigner form can be selected with fixed resonance
parameters, $S_{11}(1535)$, $S_{11}(1650)$, $P_{11}(1710)$, $P_{13}(1720)$,
$D_{13}(1520)$, $D_{13}(1700)$, $D_{15}(1675)$ and $F_{15}(1680)$~[2,3]. This model
describes almost all of the existing data of eta photoproduction on the proton from
MAMI~[4,5], ELSA~[6,7,8], GRAAL~[9,10,11] and CLAS~[12] very well. Only the target
polarization data from Bonn~[7] near threshold is in disagreement with our isobar model,
and all other existing models as well. Electroproduction data from JLab~[13,14] are also
well described and determine the transition form factor of the $S_{11}(1535)$ resonance
very precisely. This transition form factor is an exception among all other form factors,
as it drops down with $Q^2$ very slowly. In fact, it even rises against the standard
nucleon dipole form factor.

Very recently, both at GRAAL~[15] and at CB-ELSA~[16] eta photoproduction was measured on
the deuteron, allowing a separation of proton and neutron data in the range of $W_{thr.}
< W < 2$~GeV. In both experiments a bump was observed in the neutron cross sections near
$W=1675$~MeV. This bump can be reasonably well described with ETAMAID, however, only
because of a rather strong $D_{15}(1675)$ resonance with a branching ratio into the $\eta
N$ channel of $\beta_{\eta N}=17\%$. Alternatively, the bump can also be described with a
narrow $P_{11}(1675)$ resonance, following the suggestion by Polyakov~[17,18,19], who
predicted such a state as a non-strange member of the $\Theta^+$ pentaquark decuplet. In
collaboration with Alexander Fix in Mainz~[20] we have shown that such a narrow resonance
with a total width of only about $10$~MeV will give similar results as the much broader
$D_{15}$ resonance with a total width of $150$~MeV, when it is Fermi averaged in a
quasifree production process using deuteron wave functions.

Even if both pictures can lead to very similar total cross sections, in angular
distributions of cross sections and polarization observables they can lead to rather
different shapes due to interferences with different multipoles, most likely with the
dominant $S_{11}$ partial wave. A first comparison with preliminary angular distributions
from CB-ELSA at $W=1025$~MeV and $1075$~MeV leads to the finding that the conventional
$D_{15}$ model is in a qualitatively better agreement than the narrow $P_{11}$ model.

\newpage

\section{List of Participants}

\begin{itemize}
\item Vadim Baru $<$baru@itep.ru$>$, 
Institute of Theoretical and Experimental Physics, Moscow, Russia

\item Michael D\"oring $<$michael.doering@ific.uv.es$>$, Departamento de F\'\i sica Te\'orica and IFIC, Universidad de Valencia

\item Goran F\"aldt $<$goran.faldt@tsl.uu.se$>$, The Svedberg Laboratory, Uppsala, Sweden

\item Alexander Fix $<$fix@kph.uni-mainz.de$>$, Institut f\"ur Kernphysik,
Johannes Guten{-}berg-Universt\"at, Mainz, Germany

\item Humberto Garcilazo $<$humberto@Gina.esfm.ipn.mx$>$, Instituto Polit\'ecnico Nacional, Edificio 9,
07738 M\'exico D.F., Mexico

\item Ashot Gasparyan $<$gasparyn@itep.ru$>$, Institute of Theoretical and Experimental Physics,
Moscow, Russia

\item Johann Haidenbauer $<$j.haidenbauer@fz-juelich.de$>$, Institut f\"ur Kernphysik, FZ J\"ulich, Germany

\item Christoph Hanhart $<$c.hanhart@fz-juelich.de$>$, Institut f\"ur Kernphysik, FZ J\"ulich, Germany

\item Bo H\"oistad $<$bo.hoistad@tsl.uu.se$>$,  Department of
Radiation Sciences of Uppsala University,
 Uppsala, Sweden

\item Daniil Kirillov $<$da.kirillov@fz-juelich.de$>$, Institut f\"ur Kernphysik, FZ J\"ulich, Germany

\item Bernd Krusche $<$Bernd.Krusche@unibas.ch$>$, Department of Physics and Astronomy, University of Basel,
Ch-4056 Basel, Switzerland

\item Alexander Kudryavtsev $<$kudryavt@itep.ru$>$,
Institute of Theoretical and Experimental Physics, Moscow, Russia

\item Sasha Kvinikhidze $<$sasha\_kvinikhidze@hotmail.com$>$, Mathematical Institute of Georgian Academy of Sciences, Tbilisi,  Georgia

\item Timo Mersmann $<$mersmat@uni-muenster.de$>$, Institut f\"ur Kernphysik,
Universit\"at M\"unster, M\"unster, Germany

\item Ulrich Mosel $<$Ulrich.Mosel@theo.physik.uni-giessen.de$>$, Institut f\"ur Theoretische Physik, Universt\"at Giessen, Giessen, Germany

\item Pawel Moskal $<$p.moskal@fz-juelich.de$>$, Institute of Physics, Jagellonian University, and
IKP, Forschungszentrum J\"ulich

\item Jouni Niskanen $<$jouni.niskanen@helsinki.fi$>$, Department of Physical Sciences, PO Box 64, FIN-00014
University of Helsinki, Finland

\item Andreas Nogga $<$a.nogga@fz-juelich.de$>$, Institut f\"ur Kernphysik, FZ J\"ulich, Germany

\item Eulogio Oset $<$oset@ific.uv.es$>$, Departamento de F\'\i sica Te\'orica and IFIC, Universidad de Valencia

\item Teresa Pena $<$teresa@fisica.ist.utl.pt$>$,  Instituto Superior T\'ecnico,
Av. Rovisco Pais, P-1049-001 Lisboa, Portugal

\item Vitaliy Shklyar $<$Vitaliy.Shklyar@theo.physik.uni-giessen.de$>$, 
Institut f\"ur Theoreti{-}sche Physik, Universt\"at Giessen, Giessen, Germany

\item Alexander Sibirtsev $<$a.sibirtsev@fz-juelich.de$>$, Institut f\"ur Kernphysik, FZ J\"ulich, Germany

\item Aleksandr Starostin $<$starost@ucla.edu$>$, University of California, Los Angeles, CA 90095-1547, USA

\item Hans Stroeher $<$h.stroeher@fz-juelich.de$>$, Institut f\"ur Kernphysik, FZ J\"ulich, Germany

\item Alfred \v{S}varc $<$svarc@irb.hr$>$, Ru\dd er Bo\v skovi\' c Institute, Bijeni\v cka cesta 54, 10000 Zagreb, Croatia.

\item Lothar Tiator $<$tiator@kph.uni-mainz.de$>$, Institut f\"ur Kernphysik,
Johannes Guten{-}berg-Universt\"at, Mainz, Germany

\item Colin Wilkin $<$cw@hep.ucl.ac.uk$>$, Physics Department, University
College London, London WC1E 6BT, U.K.

\item Magnus Wolke $<$m.wolke@fz-juelich.de$>$, Institut f\"ur Kernphysik, FZ J\"ulich, Germany

\item Jozef Zlomanczuk $<$Jozef.Zlomanczuk@tsl.uu.se$>$, Department of
Radiation Sciences of Uppsala University,
 Uppsala, Sweden.
\end{itemize}


\begin{thebibliography}{99}
\bibitem{Krusche_03} 
B. Krusche and S. Schadmand,
  Prog.\ Part.\ Nucl.\ Phys.\  {\bf 51} (2003) 399
\bibitem{Krusche_05} B. Krusche, Prog. Part. Nucl. Phys.,
\Journal{\PPNP} {55} {46} {2005}
\bibitem{Hicks_73}
H.R. Hicks, S.R. Deans, D.T. Jocabs, P.W. Lyons, D.L. Montgomery,
\Journal{\PRD} {7} {2614} {1973}
\bibitem{Tabakin_89}
F. Tabakin, S.A. Dytman, A.S. Rosenthal,
\Journal{\em Proceedings of the Excited Baryons Conf., Troy, New York, USA 4-6
August 1988, Edts. G. Adams, N.C. Mukhopadhyay, P. Stoler, World Scientific}
{} {168} {1989}
\bibitem{Homma_88}
S. Homma et al.,
\Journal{\JPSJ} {57} {828} {1988} 
\bibitem{Price_95}
J.W. Price et al.,
\Journal{\PRC} {51} {R2283} {1995}
\bibitem{Krusche_95}
B. Krusche et al.,
\Journal{\PRL} {74} {3736} {1995}
\bibitem{Krusche_95a}
B. Krusche et al.,
\Journal{\PRL} {75} {3023} {1995}
\bibitem{Krusche_95b}
B. Krusche et al.,
\Journal{\ZPA} {351} {237} {1995}
\bibitem{Renard_02}
F. Renard et al.,
\Journal{\PLB} {528} {215} {2002}
\bibitem{Dugger_02}
M. Dugger et al.,
\Journal{\PRL} {89} {222002} {2002}
\bibitem{Crede_05}
V. Cred\'e et al.,\Journal{\PRL} {94} {012004} {2005}
\bibitem{Ajaka_98}
J. Ajaka et al.,
\Journal{\PRL} {81} {1797} {1998}
\bibitem{Bock_98}
A. Bock et al.,
\Journal{\PRL} {81} {534} {1998}
\bibitem{Schoch_95}
B. Schoch,
\Journal{\PPNP} {34} {43} {1995}
\bibitem{Thompson_01} 
R. Thompson et al.,
\Journal{\PRL} {86} {1702} {2001}
\bibitem{Armstrong_99}
C.S. Armstrong et al.,
\Journal{\PRD} {60} {052004} {1999}
\bibitem{Knochlein_95}
G. Kn{\"o}chlein, D. Drechsel, L. Tiator,
\Journal{\ZPA} {352} {327} {1995}
\bibitem{Sauermann_95}
C. Sauermann, B.L. Friman, W. N\"orenberg,
\Journal{\PLB} {341} {261} {1995}
\bibitem{Li_95a}
Zhenping Li,
\Journal{\PRD} {52} {4961} {1995}
\bibitem{Benmerrouche_96} 
M. Benmerrouche, N.C. Mukhopadhyay, J.F. Zhang,
\Journal{\PRL} {77} {4716} {1996}
\bibitem{Krusche_97}
B. Krusche, N.C. Mukhopadhyay, J.F. Zhang, M. Benmerrouche,
\Journal{\PLB} {397} {171} {1997}
\bibitem{Mathur_98}
N.C. Mukhopadhyay and N. Mathur,
\Journal{\PLB} {444} {7} {1998} 
\bibitem{Tiator_99} 
L. Tiator, D. Drechsel, G. Kn\"ochlein, C. Bennhold,
\Journal{\PRC} {60} {035210} {1999}
\bibitem{Saghai_01}
B. Saghai and Z. Li,
\Journal{\EPJA} {11} {217} {2001}
\bibitem{Chiang_02}
W.-T. Chiang, S.N. Yang, L. Tiator, D. Drechsel,
\Journal{\NPA} {700} {429} {2002}
\bibitem{Chiang_03}
W.-T. Chiang, S.N. Yang, M. Vanderhaeghen, D. Drechsel,
\Journal{\NPA} {723} {205} {2003}
\bibitem{Hyodo_03} T. Hyodo, S.I. Nam, D. Jido, A. Hosaka,
\Journal{nucl-th} {\it 0305023/} {} {2003} 
\bibitem{Liu_05}
J. Liu, J. He, and Y.B. Dong,
\Journal{\PRD} {71} {094004} {2005} 
\bibitem{Glozman_96}
L.Ya. Glozman and D.O. Riska,
\Journal{\PREP} {268} {263} {1996}
\bibitem{Glozman_96a}
L.Ya. Glozman and D.O. Riska,
\Journal{\PLB} {366} {305} {1996}
\bibitem{Kaiser_95}
N. Kaiser, P.B. Siegel, W. Weise,
\Journal{\PLB} {362} {23} {1995}
\bibitem{Kaiser_97} 
N. Kaiser, T. Waas, W. Weise,
\Journal{\NPA} {612} {297} {1997}
\bibitem{Bacci_69}
C. Bacci et al., 
\Journal{\PLB} {28} {687} {1969}
\bibitem{Krusche_95c}
B. Krusche et al.,
\Journal{\PLB} {358} {40} {1995}
\bibitem{Hoffmann-Rothe_97}
P. Hoffmann-Rothe et al.,
\Journal{\PRL} {78} {4697} {1997}
\bibitem{Weiss_03}
J. Weiss et al.,
\Journal{\EPJA} {16} {275} {2003}
\bibitem{Hejny_02}
V. Hejny et al.,
\Journal{\EPJA} {13} {493} {2002}
\bibitem{Kouznetsov_04}
V. Kouznetsov et al.,
\Journal{Proceedings of NSTAR 2004, Grenoble, France 2004} {} {197} {2004}
\bibitem{Jaegle_05}
I. Jaegle et al.,
\Journal{Proceedings of NSTAR 2005, Tallahasse, USA} {} {,in press} {2005}
\bibitem{Anderson_69}
R.L. Anderson, R. Prepost
\Journal{\PRL} {23} {46} {1969}
\bibitem{Weiss_01}
J. Weiss et al.,
\Journal{\EPJA} {11} {371} {2001}
\bibitem{Hejny_99}
V. Hejny et al.,
\Journal{\EPJA} {6} {83} {1999}
\bibitem{Pfeiffer_04}
M. Pfeiffer et al.,
\Journal{\PRL} {92} {252001} {2004}
\bibitem{Pfeiffer_05}
M. Pfeiffer et al.,
\Journal{\PRL} {94} {049102} {2005}
\bibitem{Calen_96}
H. Cal\'en et al., 
\Journal{\PLB} {366} {39} {1996} 
\bibitem{Plouin_90}
F. Plouin et al., 
\Journal{\PRL} {65} {690} {1990} 
\bibitem{Calen_98}
H. Cal\'en et al., 
\Journal{\PRL} {80} {2069} {1998} 
\bibitem{Mayer_96}
B. Mayer et al., 
\Journal{\PRC} {53} {2068} {1996} 
\bibitem{Willis_97}
N. Willis et al., 
\Journal{\PLB} {406} {14} {1997}   
\bibitem{Hibou_00}
F. Hibou et al., 
\Journal{\EPJA} {7} {537} {2000}; 
\bibitem{Bilger_02}
R. Bilger et al., 
\Journal{\PRC} {65} {044608} {2002} 
\bibitem{Chrien_88}
R.E. Chrien et al.,
\Journal{\PLB} {60} {2595} {1988}
\bibitem{Johnson_93}
J. D. Johnson et al.,\Journal{\PRC} {47} {2571} {1993}
\bibitem{Sokol_00} 
G.A. Sokol et al., 
\Journal{\it nucl-ex/} {0011005} {} {2000} 
\bibitem{Sokol_01} 
G.A. Sokol et al., 
\Journal{\it nucl-ex/} {0106005} {} {2001}
\bibitem{Nimai_95}
N.C. Mukhopadhyay, J.F. Zhang, M. Benmerrouche,
\Journal{\PLB} {364} {1} {1995} 
\bibitem{Fix_97}
A. Fix, H. Arenh{\"ovel},
\Journal{\ZPA} {359} {427} {1997}
\bibitem{Ritz_99}
F. Ritz, H. Arenh{\"o}vel,
\Journal{\PLB} {447} {15} {1999}
\bibitem{Fix_00}
A. Fix, H. Arenh{\"ovel},
\Journal{\PLB} {492} {32} {2000}
\bibitem{Fix_00a}
A. Fix, H. Arenh{\"ovel},
\Journal{\EPJA} {9} {119} {2000}
\bibitem{Ritz_01}
F. Ritz, H. Arenh{\"o}vel, 
\Journal{\PRC} {64} {034005} {2001}
\bibitem{Sibirtsev_01}
A. Sibirtsev et al.,
\Journal{\PRC} {64} {024006} {2001}
\bibitem{Fix_02} 
A. Fix, H. Arenh{\"ovel},
\Journal{\NPA} {697} {277} {2002}
\bibitem{Sibirtsev_02}
A. Sibirtsev et al.,
\Journal{\PRC} {65} {067002} {2002}
\bibitem{Fix_04} 
A. Fix, H. Arenh{\"ovel},
\Journal{\EPJA} {19} {275} {2004}
\bibitem{Fix_02a}
A. Fix, H. Arenh{\"ovel},
\Journal{\PRC} {66} {024002} {2002}
\bibitem{Fix_03} 
A. Fix, H. Arenh{\"ovel},
\Journal{\PRC} {68} {190} {2003}
\bibitem{Sibirtsev_04}
A. Sibirtsev et al.,
\Journal{\PRC} {70} {047001} {2004}
\bibitem{Hanhart_05}
C. Hanhart,
\Journal{\PRL} {2005} {94} {049101} {2005}
\bibitem{Bhalerao_85} 
R.S.Bhalerao and L.C.Liu, 
\Journal{\PRL} {54} {865} {1985} 
\bibitem{Liu_86} 
L.C.Liu and Q.Haider, 
\Journal{\PRC} {34} {1845} {1986}
\bibitem{Ueda_91} 
T.Ueda, 
\Journal{\PRL} {66} {297} {1991} 
\bibitem{Ueda_92} 
T.Ueda, 
\Journal{\PLB} {291} {228} {1992}
\bibitem{Wilkin_93} 
C.Wilkin, 
\Journal{\PRC} {47} {R938} {1993} 
\bibitem{Rakityanski_95} 
S.A.Rakityanski et al., 
\Journal{\PLB} {359} {33} {1995} 
\bibitem{Rakityanski_96} 
S.A. Rakityanski et al., 
\Journal{\PRC} {53} {R2043} {1996} 
\bibitem{Green_96} 
A.M. Green and S. Wycech, 
\Journal{\PRC} {54} {1970} {1996} 
\bibitem{Green_97} 
A.M. Green and S. Wycech, 
\Journal{\PRC} {55} {R2167} {1997}  
\bibitem{Scoccola_98} 
N.N. Scoccola, D.O. Riska, 
\Journal{\PLB} {444} {21} {1998}  
\bibitem{Green_99} 
A.M. Green and S. Wycech, 
\Journal{\PRC} {60} {35208} {1999} 
\bibitem{Shevchenko_00} 
N.V. Shevchenko et al., 
\Journal{\EPJA} {9} {143} {2000}
\bibitem{Grishina_00} 
V. Yu. Grishina et al., 
\Journal{\PLB} {475} {9} {2000}  
\bibitem{Garcilazo_01} 
H. Garcilazo and M.T. Pena, 
\Journal{\PRC} {63} {R21001} {2001} 
\bibitem{Sibirtsev_02a} 
A. Sibirtsev et al., 
\Journal{\PRC} {65} {044007} {2002}
\bibitem{Bacci_69a}
C. Bacci et al.,
\Journal{\LNC} {8} {391} {1969}
\bibitem{Roebig_96} 
M. R{\"o}big-Landau et al.,
\Journal{\PLB} {373} {45} {1996}
\bibitem{Yorita_00} 
T. Yorita et al.
\Journal{\PLB} {476} {226} {200}
\bibitem{Yamazaki_00} 
H. Yamazaki et al.,
\Journal{\NPA} {670} {202c} {2000} 
\bibitem{Kinoshita_05}
T. Kinoshita et al.,
\Journal{nucl-ex/} {0509022} {} {2005}
\bibitem{Mertens_06}
T. Mertens et al.,
\Journal{priv. com.} {} {} {2006}
\bibitem{Krusche_04}
B. Krusche et al.,
\Journal{\EPJA} {22} {277} {2004} 
\bibitem{Krusche_04a}
B. Krusche et al.,
\Journal{\EPJA} {22} {347} {2004}
\bibitem{Carrasco_93} 
R.C. Carrasco,
\Journal{\PRC} {48} {2333} {1993}
\bibitem{Hombach_95} 
A. Hombach et al.,
\Journal{ZPA} {352} {223} {1995}
\bibitem{Lehr_00}
J. Lehr, M. Effenberger, U. Mosel,
\Journal{\NPA} {671} {503} {2000}  
\bibitem{Lehr_01} 
J. Lehr and U. Mosel,
\Journal{\PRC} {64} {042202} {2001}
\bibitem{Lehr_03} 
J. Lehr, M. Post and U. Mosel,
\Journal{\PRC} {68} {044601} {2003}
\bibitem{Post_04}
M. Post, J. Lehr, U. Mosel,
\Journal{\NPA} {741} {81} {2004}
\end{thebibliography}

\begin{thebibliography}{99}
\bibitem{a1}  A.~Starostin {\it et al.}, Phys.\ Rev.\ Lett. {\bf 85}, 5539 (2000). 
\bibitem{a2}  A.~Starostin {\it et al.}, Phys.\ Rev.\ C {\bf 66}, 055205 (2002). 
\bibitem{a3}  S.~Prakhov {\it et al.}, Phys.\ Rev.\ C {\bf 72}, 015203 (2005). 
\end{thebibliography}

\begin{thebibliography}{99}
\bibitem{b1} A. Sibirtsev {\it et~al.}, Phys. Rev. C 70, 047001 (2004).
\bibitem{b2} A. Sibirtsev {\it et~al.}, {Eur. Phys. J.} {A 22} 495 (2004).
\bibitem{b3} M. Pfeiffer {\it et~al.}, {Phys. Rev. Lett.} {92}, 252001 (2004).
\bibitem{b4} J. Berger {\it et~al.}, {Phys. Rev. Lett.} {61}, 919 (1988).
\bibitem{b5} B. Mayer {\it et~al.}, {Phys. Rev. C} {53}, 2068 (1996).
\bibitem{b6} A. Khoukaz, T. Mersmann, COSY Proposal 137 (2004).
\end{thebibliography}

\begin{thebibliography}{99}
\bibitem{c[1]} B.J. Roy et al; Phys. Lett. B 550(2002) 47
\bibitem{c[2]} B.J. Roy et al; BARC external report no. BARC/2000/E/004,\\
Bhabha Atomic Research Centre, Mumbai-400085, India
\bibitem{c[3]} L.C. Liu; private communication
\bibitem{c[4]} R.S. Bhalerao et al; Phys. Rev. Lett. 54 (1985) 865
\bibitem{c[5]} C. Wilkin, Phys. Rev. C 47 (1993) R398
\bibitem{c[6]} B.J. Roy et al; Cosy proposal No.50.3, COSY, J\"ulich
\end{thebibliography}

\begin{thebibliography}{99}
\bibitem{proposal}
J. Smyrski, P. Klaja, P. Moskal, COSY Proposal {\bf No. 160}\\
available at: 
{\small{http://www.fz-juelich.de/ikp/publications/List$\_$of$\_$all$\_$COSY-Proposals.shtml}}

\bibitem{eta05}
J.~Smyrski et al.,
Acta Phys. Slovaca {\bf 56} (2006) 213 [nucl-ex/0603023]

\bibitem {Hai} Q. Haider and L. C. liu, Phys. Lett. {\bf B172} (1986) 257

\bibitem {Bas} S. D. Bass and A. W. Thomas, Phys. Lett. {\bf B634} (2006) 368

\bibitem {Cal} H. Cal\'en et al., {\bf 80} (1998) 2069

\bibitem {Wil} C. Wilkin, Phys. Rev. {\bf C47} (1993) R938

\bibitem {Wis} N. Willis et al., Phys. Lett. {\bf B406} (1997) 14 

\bibitem {Wyc} S. Wycech, A. M. Green and J. A. Niskanen, Phys. Rev. {\bf C52} (1995) 544

\bibitem {Sco} N. N. Scoccola and D. O. Riska, Phys. Lett. {\bf B444} (1998) 21

\bibitem {Pfe} M. Pfeiffer et al., Phys. Rev. Lett. {\bf 92} (2004) 252001

\bibitem {christoph} C. Hanhart, e-Print Archive: hep-ph/0408204

\bibitem {Fet} J. G. Fetkovich et al., Phys. Rev. {\bf D6} (1972) 3069
\end{thebibliography}

\begin{thebibliography}{99}
\bibitem{1d} F. Plouin et al., \emph{Nucl. Phys}. A 302, 413 (1978). 

\bibitem{2d} F. Plouin, P. Fleury, and C. Wilkin, \emph{Phys. Rev. Lett}. 65,
690 (1990). 

\bibitem{3d} H. Cal\'en et al., \emph{Phys. Rev. Lett}. 79, 2642 (1997). 

\bibitem{4d} H. Cal\'en et al., \emph{Phys. Rev. Lett}. 80, 2069 (1998). 

\bibitem{5d} J. Berger et al., \emph{Phys. Rev. Lett.} 61, 919 (1988). 

\bibitem{6d} B. Mayer et al, \emph{Phys. Rev}. C 53, 2068 (1996). 

\bibitem{7d} R. Frascaria et al.,\emph{ Phys. Rev}. C 50, R537 (1994). 

\bibitem{8d} N. Willis et al., \emph{Phys. Lett}. B 406, 143 (1997). 

\bibitem{9d} T. Ueda, Phys. \emph{Rev. Lett.} 66, 297 (1991); Physica Scripta
48, 68 (1993). 

\bibitem{10d} S. Wycech, A.M. Green, and J.A. Niskanen, \emph{Phys. Rev.} C 52,
544 (1995). 

\bibitem{11d} S.A. Rakityansky, S.A. Sofianos, W. Sandhas, and V.B. Belyaev,
\emph{Phys. Lettd} B 359, 33 (1995). 

\bibitem{12d} V.B. Belyaev, S.A. Rakityansky, S.A. Sofianos, M. Braun, and W.
Sandhas, \emph{Few. Body. Syst. Suppl}. 8, 309 (1995). 

\bibitem{13d} S.A. Rakityansky, S.A. Sofianos, M. Braun, V.B. Belyaev and W.
Sandhas, \emph{Phys. Rev. }C 53, R2043 (1996). 

\bibitem{14d} A. Fix and H. Arenhvel, \emph{Phys. Rev.} C 66, 024002 (2002). 

\bibitem{15d} C. Wilkin, Phys. \emph{Rev. C}. 47, R938 (1993). 

\bibitem{16d} Q. Haider and L.C. Liu, \emph{Phys. Lett}. B 172, 257 (1986). 

\bibitem{17d} L.C. Liu and Q. Haider, \emph{Phys. Rev.} C 34, 1845 (1986). 

\bibitem{18d} H.C. Chiang, E. Oset, and L.C. Liu, \emph{Phys. Rev}. C 44, 738
(1991). 

\bibitem{19d} Q. Haider and L.C. Liu, \emph{Phys. Rev}. C 66, 045208 (2002). 

\bibitem{20d} M. Pfeiffer et al., \emph{Phys. Rev. Lett}. 92 (2004) 252001. 

\bibitem{21d} C. Hanhart, \emph{arXiv:hep-ph/0408204 v1} 18 Aug 2004. 22.
\end{thebibliography}

\begin{thebibliography}{99}
\bibitem{1e} A.N. Kvinikhidze, B. Blankleider. Phys.Lett.B307:7-12 (1993)
\bibitem{2e} A.T. Stelbovics, M. Stingl, Nucl.Phys. A294, 391 (1978)
\end{thebibliography}

\begin{thebibliography}{99}
\bibitem{Gre84} W. Grein, A. K\"{o}nig, P. Kroll, M.P. Locher and A. \v{S}varc, Annals of Physics {\bf 153} (1984) 
301.
\bibitem{Bat94} M. Batini\'{c}, T.-S.H. Lee, M.P. Locher, Yang Lu and A. \v{S}varc, Phys. Rev. C {\bf 50} (1994) 
1300.
\bibitem{Gar02} H. Garcilazo and M.T. Pe\~{n}a, Phys. Rev. C {\bf 66} (2002) 034606.
\bibitem{Gar05} H. Garcilazo and M.T. Pe\~{n}a, Phys. Rev. C {\bf 72} (2005) 014003.
\bibitem{Bat98} M. Batini\'{c}, I. \v{S}laus, A. \v{S}varc and B.M.K. Nefkens,
Phys. Rev. {\bf C51}, 2310 (1995); 
 M. Batini\'{c}, I. Dadi\'{c}, I. \v{S}laus, A. \v{S}varc, B.M.K. Nefkens and
T.S.-H. Lee, Physica Scripta {\bf 58}, 
15 (1998).
\bibitem{Sva06} http://www.fz-juelich.de/ikp/etanucleus/
\bibitem{Sva06a} e-mail:alfred.svarc@irb.hr 
\end{thebibliography}

\begin{thebibliography}{99}
\bibitem{f1} C. Wilkin, Phys. Rev. C.  {\bf 47}, R938 (1993).
\bibitem{f2} S. Wycech, A.M. Green and J.A. Niskanen, Phys. Rev.
    C {\bf 52}, 544 (1995).
\bibitem{f3} S.A. Rakityansky, S.A. Sofianos, M. Braun, V.B. Belyaev
    and W. Sandhas, Phys. Rev. C {\bf 53}, R2043 (1996).
\bibitem{f4} A. Fix and H. Arenh\"ovel, Phys. Rev. C {\bf 66}, 024002 (2002).
\bibitem{f5} Q. Haider and L.C. Liu, Phys. Rev. C {\bf 66}, 045208 (2002).
\bibitem{f6}  A. Sibirtsev, J. Haidenbauer, C. Hanhart, and J. A. Niskanen,
         arXiv:nucl-th/0310079,  Eur. Phys. Journal A {\bf 22}, 495
         (2004).
\bibitem{f7} A.M. Green and S. Wycech, Phys. Rev. C {\bf 68}, 061601(R)
(2003).
\bibitem{f8} J. A. Niskanen, arXiv:nucl-th/0508021.

\end{thebibliography}

\begin{thebibliography}{99}
\bibitem{g1} H. Cal\'en {\it et al.},   
               Phys. Rev. Lett. {\bf 79}, 2642 (1997).


\bibitem{g2} H. Cal\'en {\it et al.},   
               Phys. Rev. Lett. {\bf 80}, 2069 (1998).


\bibitem{g3} H. Garcilazo and M. T, Pe\~na,
Phys. Rev. C {\bf 66}, 034006 (2002).


\bibitem{g4} H. Garcilazo and M. T. Pe\~na,
Phys. Rev. C {\bf 72}, 014003 (2005).


\bibitem{g5} H. Garcilazo and M. T.
Pe\~na, Phys. Rev. C {\bf 63}, 021001 (2001).


\bibitem{g6} H. Garcilazo,
Phys. Rev. C {\bf 67}, 067001 (2003).


\bibitem{g7} H. Garcilazo, 
Phys. Rev. C {\bf 71}, 048201 (2005).


\bibitem{g8} C. D. Frogatt and J. L. Petersen, Nucl. Phys.
             {\bf B129}, 89 (1977).


\bibitem{g9} A. Sibirtsev, S. Schneider, Ch. Elster, J. Haidenbauer, 
               S. Krewald, and J. Speth, Phys. Rev. C {\bf 65},
               044007 (2002).


\bibitem{g10} M. Batini\'c, I. \v Slaus, A. \v Svarc, and B. M. K.  
               Nefkens, Phys. Rev. C {\bf 51}, 2310 (1995);
               M. Batini\'c, I. Dadi\'c, I. \v Slaus, A. \v Svarc, B. M. K.  
               Nefkens, and T.-S. H. Lee, Physica Scripta {\bf 58}, 15
               (1998). 


\bibitem{g11} A. M. Green and S. Wycech, Phys. Rev. C {\bf 55},
               R2167 (1997).


\bibitem{g12} A. M. Green and S. Wycech, Phys. Rev. C {\bf 60},
               035208 (1999).
\end{thebibliography}

\begin{thebibliography}{99}
\bibitem{Oset:1997it}
E.~Oset and A.~Ramos,
Nucl.\ Phys.\ A {\bf 635} (1998) 99
[arXiv:nucl-th/9711022].

\bibitem{Oset:2001cn}
E.~Oset, A.~Ramos and C.~Bennhold,
Phys.\ Lett.\ B {\bf 527} (2002) 99
[Erratum-ibid.\ B {\bf 530} (2002) 260]
[arXiv:nucl-th/0109006].

\bibitem{Inoue:2001ip}
T.~Inoue, E.~Oset and M.~J.~Vicente Vacas,
strangeness
Phys.\ Rev.\ C {\bf 65} (2002) 035204
[arXiv:hep-ph/0110333].

\bibitem{Bai:2001ua}
J.~Z.~Bai {\it et al.}  [BES Collaboration],
Phys.\ Lett.\ B {\bf 510} (2001) 75
[arXiv:hep-ex/0105011].

\bibitem{Inoue:2002xw}
T.~Inoue and E.~Oset,
Nucl.\ Phys.\ A {\bf 710} (2002) 354
[arXiv:hep-ph/0205028].

\bibitem{Post:2003hu}
M.~Post, S.~Leupold and U.~Mosel,
Nucl.\ Phys.\ A {\bf 741} (2004) 81
[arXiv:nucl-th/0309085].


\bibitem{Garcia-Recio:2002cu}
C.~Garcia-Recio, J.~Nieves, T.~Inoue and E.~Oset,
Phys.\ Lett.\ B {\bf 550} (2002) 47
[arXiv:nucl-th/0206024].

\bibitem{Nagahiro:2003iv}
H.~Nagahiro, D.~Jido and S.~Hirenzaki,
Phys.\ Rev.\ C {\bf 68} (2003) 035205
[arXiv:nucl-th/0304068].


\end{thebibliography}

\begin{thebibliography}{99}

\bibitem{Lehr} J.~Lehr, M.~Post and U.~Mosel,
Phys.\ Rev.\ C {\bf 68} (2003) 044601 [arXiv:nucl-th/0306024].

\bibitem{Post} M.~Post, S.~Leupold and U.~Mosel,
Nucl.\ Phys.\ A {\bf 741} (2004) 81 [arXiv:nucl-th/0309085].

\bibitem{GiBUU} For details see:
http://tp8.physik.uni-giessen.de:8080/GiBUU/

\bibitem{Hombach}
A.~Hombach, A.~Engel, S.~Teis and U.~Mosel,
Z.\ Phys.\ A {\bf 352} (1995) 223 [arXiv:nucl-th/9411025].

\bibitem{Effenberger}
M.~Effenberger and A.~Sibirtsev,
  from
Nucl.\ Phys.\ A {\bf 632} (1998) 99 [arXiv:nucl-th/9710054].

\bibitem{Lehrelectro}
J.~Lehr and U.~Mosel,
Phys.\ Rev.\ C {\bf 68} (2003) 044603 [arXiv:nucl-th/0307009].

\bibitem{Lehr1}J.~Lehr, M.~Effenberger and U.~Mosel,
Nucl.\ Phys.\ A {\bf 671} (2000) 503 [arXiv:nucl-th/9907091].




\end{thebibliography}

\vspace{0.5cm}

\begin{thebibliography}{99}
\bibitem{h[1]} T.~Feuster and U.~Mosel,
 Phys. Rev. {\bf C58}, 457 (1998).

\bibitem{h[2]} T.~Feuster and U.~Mosel,
 Phys. Rev. {\bf C59}, 460 (1999).

\bibitem{h[3]} G.~Penner and U.~Mosel,
 Phys. Rev. {\bf C66}, 055211 (2002).

\bibitem{h[4]} G.~Penner and U.~Mosel,
 Phys. Rev. {\bf C66}, 055212 (2002).

\bibitem{h[5]} V.~Shklyar, G.~Penner, and U.~Mosel,
 Eur. Phys. J. {\bf A21}, 445 (2004).

\bibitem{h[6]} V.~Shklyar, H.~Lenske, U.~Mosel, and G.~Penner,
 Phys. Rev. {\bf C71}, 055206 (2005).

\bibitem{h[7]} V.~Shklyar, H.~Lenske, and U.~Mosel,
 Phys. Rev. {\bf C72}, 015210 (2005).

\bibitem{h[8]} A.~M. Gasparyan, J.~Haidenbauer, C.~Hanhart, and J.~Speth,
 Phys. Rev. {\bf C68}, 045207 (2003).

\bibitem{h[9]} GRAAL, F.~Renard {\em et~al.},
 Phys. Lett. {\bf B528}, 215 (2002).

\bibitem{h[10]} CLAS, M.~Dugger {\em et~al.},
 Phys. Rev. Lett. {\bf 89}, 222002 (2002).

\bibitem{h[11]} CB-ELSA, V.~Crede {\em et~al.},
 Phys. Rev. Lett. {\bf 94}, 012004 (2005).

\bibitem{h[12]} B.~Krusche {\em et~al.},
 Phys. Lett. {\bf B358}, 40 (1995).

\bibitem{h[13]} B.~Krusche {\em et~al.},
 Phys. Rev. Lett. {\bf 74}, 3736 (1995).

\bibitem{h[14]} J.~Ajaka {\em et~al.},
 Phys. Rev. Lett. {\bf 81}, 1797 (1998).

\bibitem{h[15]} W.-T. Chiang, S.~N. Yang, L.~Tiator, M.~Vanderhaeghen, and D.~Drechsel,
 Phys. Rev. {\bf C68}, 045202 (2003).

\bibitem{h[16]} W.-T. Chiang, S.-N. Yang, L.~Tiator, and D.~Drechsel,
 Nucl. Phys. {\bf A700}, 429 (2002).

\bibitem{h[17]} R.~A. Arndt, I.~I. Strakovsky, and R.~L. Workman,
 Phys. Rev. {\bf C53}, 430 (1996).
\end{thebibliography}

\begin{thebibliography}{99}
\bibitem{i1} 
A. Green and S. Wycech, Phys. Rev. C 60, 035208 (1999); nucl-th/9905011
\bibitem{i2}
V. Shklyar, H. Lenske, U. Mosel, Phys. Rev. C 72, 015210 (2005); 
nucl-th/0505010
\bibitem{i3}
A.M. Gasparyan, J. Haidenbauer, C. Hanhart, and J. Speth, Phys. Rev.
C 68, 045207 (2003); nucl-th/0307072
\bibitem{i4}
N. V. Shevchenko, S. A. Rakityansky, S. A. Sofianos, V. B. Belyaev,
and W. Sandhas, Phys. Rev. C 58, R3055 (1998) 
\bibitem{i5}
H. Garcilazo and M. T. Pe\~na, Phys. Rev. C 61, 064010 (2001) 
\bibitem{i6}
A. Fix and H. Arenh\"ovel, Nucl. Phys. A 697, 277 (2002)
\bibitem{i7}
A. Fix and H. Arenh\"ovel, Phys. Rev. C 66, 024002 (2003);
Phys. Rev. C, 68 044002 (2003) 
\bibitem{i8}
N. Kaiser, T. Waas, and W. Weise, Nucl. Phys. A 612, 297 (1997); 
hep-ph/9607459
\bibitem{i9}
T. Inoue and E. Oset, Nucl. Phys. A 710, 354 (2002);
hep-ph/0205028
\end{thebibliography}

\begin{thebibliography}{99}
\bibitem{Kaiser:1995cy}
N.~Kaiser, P.~B.~Siegel and W.~Weise,
Phys.\ Lett.\ B {\bf 362} (1995) 23.
\bibitem{Kaiser:1996js}
N.~Kaiser, T.~Waas and W.~Weise,
Nucl.\ Phys.\ A {\bf 612} (1997) 297
\bibitem{kaon}
E.~Oset and A.~Ramos,
Nucl.\ Phys.\ A {\bf 635} (1998) 99 

\bibitem{Nacher:1999vg} J.~C.~Nacher, A.~Parreno, E.~Oset, A.~Ramos, A.~Hosaka
and M.~Oka,
Nucl.\ Phys.\ A {\bf 678} (2000) 187
\bibitem{Oller:2000fj} J.~A.~Oller and U.~G.~Mei\ss ner,
Phys.\ Lett.\ B {\bf 500} (2001) 263 
\bibitem{Jido:2003cb} D.~Jido, J.~A.~Oller, E.~Oset, A.~Ramos and
U.~G.~Mei\ss ner,
Nucl.\ Phys.\ A {\bf 725}
(2003) 181 
\bibitem{Garcia-Recio:2003ks} C.~Garcia-Recio, M.~F.~M.~Lutz and J.~Nieves,
Phys.\ Lett.\ B {\bf 582} (2004) 49 
\bibitem{Inoue:2001ip2} T.~Inoue, E.~Oset and M.~J.~Vicente Vacas,
Phys.\ Rev.\ C {\bf 65} (2002) 035204 
\bibitem{Isgur:1978xj} N.~Isgur and G.~Karl,
Phys.\ Rev.\ D {\bf 18} (1978) 4187.
\bibitem{Capstick:1993kb} S.~Capstick and W.~Roberts,
Phys.\ Rev.\ D {\bf 49} (1994) 4570 
\bibitem{Chiu:2005zc} T.~W.~Chiu and T.~H.~Hsieh,
arXiv:hep-lat/0501021.
\bibitem{Nakajima:2001js} N.~Nakajima, H.~Matsufuru, Y.~Nemoto and H.~Suganuma,
arXiv:hep-lat/0204014.
\bibitem{Baru:2003qq}
V.~Baru, J.~Haidenbauer, C.~Hanhart, Y.~Kalashnikova and A.~Kudryavtsev,
Phys.\ Lett.\ B {\bf 586}, 53 (2004)
\bibitem{Sibirtsev:2001hz}
A.~Sibirtsev, S.~Schneider, C.~Elster, J.~Haidenbauer, S.~Krewald and J.~Speth,
Phys.\ Rev.\ C {\bf 65}, 044007 (2002)
\bibitem{Doring:2005bx}
M.~Doring, E.~Oset and D.~Strottman,
Phys.\ Rev.\ C {\bf 73}, 045209 (2006)
\bibitem{Doring:2006pt}
M.~Doring, E.~Oset and D.~Strottman,
arXiv:nucl-th/0602055, accepted for publication in Phys. Lett. {\bf B}.
\bibitem{Kolomeitsev:2003kt}
E.~E.~Kolomeitsev and M.~F.~M.~Lutz,
Phys.\ Lett.\ B {\bf 585}, 243 (2004)
\bibitem{Sarkar:2004jh}
S.~Sarkar, E.~Oset and M.~J.~Vicente Vacas,
Nucl.\ Phys.\ A {\bf 750}, 294 (2005)
\bibitem{nanovanstar}
M. Nanova at the ''International Workshop On The Physics Of Excited Baryons (NSTAR 05)'',
10-15 Oct 2005, Tallahassee, Florida

\end{thebibliography}

\begin{thebibliography}{99}
\bibitem{j1} A.~M.~Gasparyan, J.~Haidenbauer, C.~Hanhart and J.~Speth,
  Phys.\ Rev.\ C {\bf 68}, 045207 (2003).
\bibitem{j2} K.~Tsushima, S.~W.~Huang and A.~Faessler,
  Phys.\ Lett.\ B {\bf 337}, 245 (1994).
\bibitem{j3} G. Penner  and U. Mosel, Phys. Rev.C65 055202(2002).
\bibitem{j4} M.~F.~M.~Lutz, G.~Wolf and B.~Friman,
  Nucl.\ Phys.\ A {\bf 706}, 431 (2002).
\bibitem{j5} N.~Kaiser, P.~B.~Siegel and W.~Weise,
  Phys.\ Lett.\ B {\bf 362}, 23 (1995).
\bibitem{j6} T.~Inoue, E.~Oset and M.~J.~Vicente Vacas,
  Phys.\ Rev.\ C {\bf 65}, 035204 (2002).
\bibitem{j7} E.~E.~Kolomeitsev and M.~F.~M.~Lutz,
  Phys.\ Lett.\ B {\bf 585}, 243 (2004).
\bibitem{j8} R.A. Arndt, I.I. Strakovsky, R.L. Workman and M.M. Pavan,
 Phys.\ Rev.\ C {\bf 52}, 2120 (1995).
\bibitem{j9}  R.M. Brown et al., Nucl.\ Phys.\ B {\bf 153}, 89 (1979).
\bibitem{j10} F. Bulos et al.,  Phys.\ Rev.\ Lett. {\bf 13}, 486 (1964).
\bibitem{j11} F. Bulos et al.,  Phys.\ Rev. {\bf 187}, 1827 (1969).
\bibitem{j12} W. Deinet et al.,  Nucl.\ Phys.\ B {\bf 11}, 495 (1969).
\bibitem{j13} B.W. Richards et al.,  Phys.\ Rev.\ D {\bf 1}, 10 (1970).
\end{thebibliography}

\begin{thebibliography}{99}
\bibitem{k1} H. Garcilazo and M. T, Pe\~na,
Phys. Rev. C {\bf 66}, 034006 (2002).


\bibitem{k2} H. Garcilazo and M. T. Pe\~na,
Phys. Rev. C {\bf 72}, 014003 (2005).


\bibitem{k3} C. Hanhart, J. Haidenbauer, O. Krehl, J. Speth,
Phys. Lett. B{\bf 444} 25 (1998).


\bibitem{k4} M. Batini\'c, I. \v Slaus, A. \v Svarc, and B. M. K.  
               Nefkens, Phys. Rev. C {\bf 51}, 2310 (1995);
               M. Batini\'c, I. Dadi\'c, I. \v Slaus, A. \v Svarc, B. M. K.  
               Nefkens, and T.-S. H. Lee, Physica Scripta {\bf 58}, 15
               (1998). 
	       

\bibitem{k5} A. M. Green and S. Wycech, Phys. Rev. C {\bf 55},
               R2167 (1997).


\bibitem{k6} A. M. Green and S. Wycech, Phys. Rev. C {\bf 60},
               035208 (1999).
\end{thebibliography}

\begin{thebibliography}{99}
\bibitem{l1} http://www.kph.uni-mainz.de/MAID/
\bibitem{l2} W.-T. Chiang, S.N. Yang, L. Tiator and D. Drechsel, Nucl. Phys. A 700, 429 (2002).
\bibitem{l3} W.-T. Chiang, S.N. Yang, L. Tiator, M. Vanderhaeghen and D. Drechsel,
Phys. Rev. C 68, 045202 (2003).
\bibitem{l4} B.~Krusche {\it et~al.}, {Phys. Rev. Lett.} {74}, 3736 (1995).
\bibitem{l5} B.~Krusche {\it et~al.}, {Phys. Lett.} { B 358}, 40 (1995).
\bibitem{l6} J.W. Price {\it et~al.}, Phys. Rev. C 51, 2283 (1995).
\bibitem{l7} A. Bock {\it et~al.}, Phys. Rev. Lett. 81, 534 (1998).
\bibitem{l8} V.~Crede {\it et~al.}, {Phys. Rev. Lett.} {94}, 012004 (2005).
\bibitem{l9} J. Ajaka {\it et~al.}, Phys. Rev. Lett. 81, 1797 (1998).
\bibitem{l10} V. Kouznetsov, SAID database, GRAAL, 2001.
\bibitem{l11} F.~Renard {\it et~al.}, Phys. Lett. B 528, 215 (2002).
\bibitem{l12} M. Dugger {\it et~al.}, {Phys. Rev. Lett.} {89}, 222002 (2002); {89}, 249904(E) (2002).
\bibitem{l13} C.S. Armstrong {\it et~al.}, {Phys. Rev.} {D 60}, 052004 (1999).
\bibitem{l14} R. Thompson {\it et~al.}, {Phys. Rev. Lett.} {86}, 1702 (2001).
\bibitem{l15} V. Kouznetsov (for the GRAAL collaboration),
Proc. of the NSTAR2005 Workshop, Tallahassee, FL, World Scientific, 2006.
\bibitem{l16} I. Jaegle (for the CB-ELSA collaboration),
Proc. of the NSTAR2005 Workshop, Tallahassee, FL, World Scientific, 2006.
\bibitem{l17} D. Diakonov, V. Petrov, and M.V. Polyakov, Z. Phys. A 359, 305 (1997).
\bibitem{l18} M.V. Polyakov and A. Rathke, Eur. Phy. J A 18, 691 (2003).
\bibitem{l19} Ya. Azimov, V. Kouznetsov, M.V. Polyakov and I. Strakovsky, Eur. Phy. J A 25, 325 (2005).
\bibitem{l20} A. Fix and H. Arenh\"ovel, Z. Phys. A 359, 427 (1997) and private communication.
\end{thebibliography}
\end{document}